# A Lightweight Space-based Solar Power Generation and Transmission Satellite

Behrooz Abiri*, Manan Arya, Florian Bohn, Austin Fikes, Matan Gal-Katziri, Eleftherios Gdoutos, Ashish Goel, Pilar Espinet Gonzalez, Michael Kelzenberg, Nicolas Lee, Michael A. Marshall, Tatiana Roy, Fabien Royer, Emily C. Warmann, Nina Vaidya, Tatiana Vinogradova, Richard Madonna†, Harry Atwater, Ali Hajimiri, Sergio Pellegrino

*Abstract*— **We propose a novel design for a lightweight, high-performance space-based solar power array combined with power beaming capability for operation in geosynchronous orbit and transmission of power to Earth. We use a modular configuration of small, repeatable unit cells, called tiles, that each individually perform power collection, conversion, and transmission. Sunlight is collected via lightweight parabolic concentrators and converted to DC electric power with high efficiency III-V photovoltaics. Several CMOS integrated circuits within each tile generates and controls the phase of multiple independently-controlled microwave sources using the DC power. These sources are coupled to multiple radiating antennas which act as elements of a large phased array to beam the RF power to Earth. The power is sent to Earth at a frequency chosen in the range of 1-10 GHz and collected with ground-based rectennas at a local intensity no larger than ambient sunlight. We achieve significantly reduced mass compared to previous designs by taking advantage of solar concentration, current CMOS integrated circuit technology, and ultralight structural elements. Of note, the resulting satellite has no movable parts once it is fully deployed and all beam steering is done electronically. Our design is safe, scalable, and able to be deployed and tested with progressively larger configurations starting with a single unit cell that could fit on a cube satellite. The design reported on here has an areal mass density of 160 g/m$^2$ and an end-to-end efficiency of 7-14%. We believe this is a significant step forward to the realization of space-based solar power, a concept once of science fiction.**

## I. INTRODUCTION

THE concept of collecting solar power in space and transmitting it to Earth using microwaves was first published in 1941 in a science fiction short story [1] and rediscovered in 1968 [2]. Locating photovoltaics in space avoids the major disadvantages of terrestrial solar energy collection such as intermittent availability (i.e. day-night cycle) and influence by changing weather conditions. Clean, renewable power could be continuously available and potentially sent to any location on Earth. Significantly more power can be collected in space than on

This work was in part supported by Caltech Provost Funds, the Northrop Grumman-Caltech Space Solar Power Initiative, and the Caltech Space Solar Power Project.

*The Authors are listed in alphabetical order, the principal investigators, H.A., A.H., and S.P. are also listed in alphabetical order.

†Dr. Richard Madonna is the corresponding author (rmadonna@caltech.edu)

Austin Fikes and Ali Hajimiri are with Department of Electrical Engineering (EE) at California Institute of Technology (Caltech), Pasadena CA. Matan Gal-Katziri, Behrooz Abiri, and Florian Bohn were formerly with the EE department at Caltech. Eleftherios Gdoutos, Michael A. Marshall and Sergio Pellegrino are with GALCIT at Caltech, Pasadena, CA. Fabien Royer and Nicolas Lee were formerly with GALCIT at Caltech. Michael Kelzenberg, Emily C. Warmann, and Harry Atwater are with the Department of Applied Physics (APh) at Caltech. Pilar Espinet Gonzalez, Nina Vaidya, and Tatiana Roy were formerly with the Department of Applied Physics at Caltech. Richard Madonna is with Caltech, Pasadena, CA.

Earth due to constant direct access to the sun and the absence of losses due to reflection and absorption of solar energy by the Earth's atmosphere. Despite these advantages, periodic evaluations of the concept for technical feasibility have all concluded that current launch costs make the full realization too expensive [3]. This is in part due to the reliance on relatively heavy off-the-shelf photovoltaic and power components that need to be launched into space, typically in numerous launches. While launch costs have decreased somewhat in recent decades [4] the cost of lifting a large-scale array to a geosynchronous orbit (GEO) for near-continuous power generation remains impractical. A more recent development effort has proposed a more lightweight design [5], but this concept has a relatively low overall efficiency (~1%) due to the use of low efficiency (~2%) solar cells, non-complete photovoltaic aperture coverage, and failure to account for DC to RF power conversion. Many of the previous designs also require assembly in space [6], either by humans or robots, significantly increasing the cost and complexity of deployment. To date, a practical design of a high performance, deployable, lightweight space-based solar power system compatible with current launch costs has yet to be presented. Here, we describe an alternate approach to space-based solar power focused on designing an integrated, lightweight, high-performance system employing numerous innovations to meet the demands of current launch costs. To do this, we leverage recent developments in photovoltaic and microelectronic technologies that allow for the design of much more lightweight, compact versions of all the necessary components for DC generation, DC to RF conversion, and RF transmission of solar power in space. In taking this approach, we trade the rigid, heavy structures used in other approaches to produce a physically and electrically flat RF aperture, for ultra-light weight structures requiring in-situ shape metrology to enable real-time adjustments to the phases of the RF emitters which produces an electrically flat surface. We believe our approach is a significant step towards realization of the long elusive space-based solar power concept.

## II. Our General Proposed Concept

Our design approach centers on a modular, integrated power generation and transmission unit that can be constructed with extremely low mass and assembled into larger area arrays. Figure 1 shows a simplified schematic of a single lightweight unit cell, called a tile, that can be repeated in an array to form a large-area space power station. This single tile contains photovoltaic devices with lightweight solar concentrating elements, a microwave frequency antenna and one or more integrated circuits to convert the DC generated power from the photovoltaics to an RF signal for antenna transmission. All the necessary components for power generation and transmission are contained in this single tile that can be repeated to collect sunlight over a larger area. The tile's electronic components are structurally supported by lightweight polyimide films and carbon fiber frames and springs. These tiles can be easily flattened and unfolded for deployment. We term a connected assembly of these tiles a "module," which can be combined with other modules to form an arbitrarily large space-based power generating station. The modules fold into compact packages for launch and deploy autonomously in orbit. The complete power station consists of an array of these modules flying in compact formation and maintaining position and attitude via module-based thrusters. The microwave transmission of each antenna across the station is time and phased controlled to create a beam-forming phased array. The transmitted power is collected on Earth by a ground-based array of rectennas. Figure **2** is a conceptual drawing of the power station showing the organizational hierarchy.

The modular approach presents two key advantages. First, the integration of solar and RF energy conversion in one tile removes the need for a power distribution network throughout the array. Eliminating the power bus reduces the weight and complexity of the overall system and improves robustness by ensuring that one tile's failure does not impact other parts of the array. A second advantage is the inherent scalability of the system. The specific dimensions and number of modules and the extent of the array are not dependent on the basic tile design, and therefore the space power station design can easily be adapted for various applications. In addition, an existing array can be expanded with the addition of modules over time without interrupting or degrading its performance.

We now examine in detail the three specific areas of study where we have focused our efforts for design of our space-based solar power array followed by additional discussion of more general and integrated system considerations.

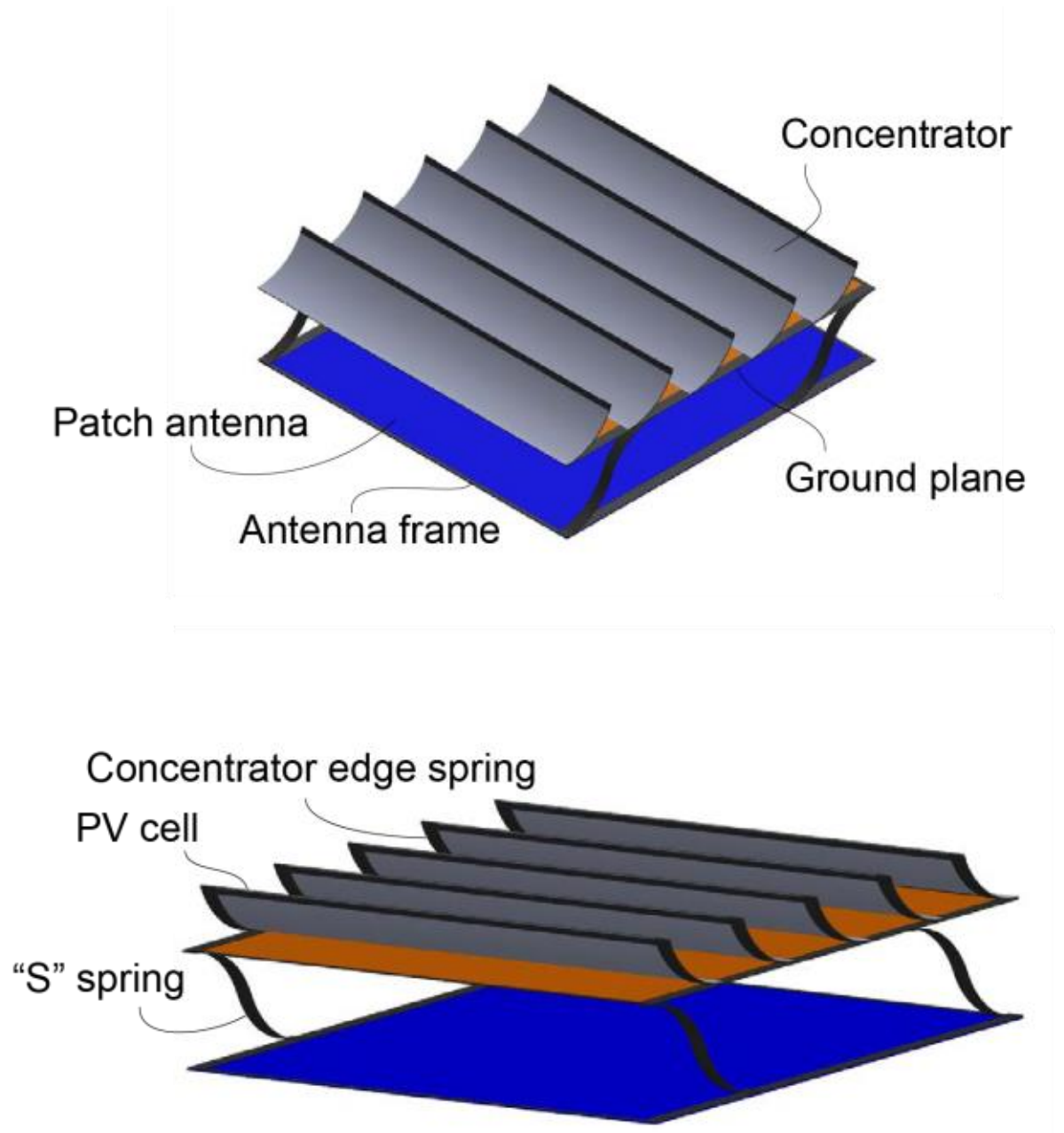

Figure 1**:** Schematic of general modular tile architecture.

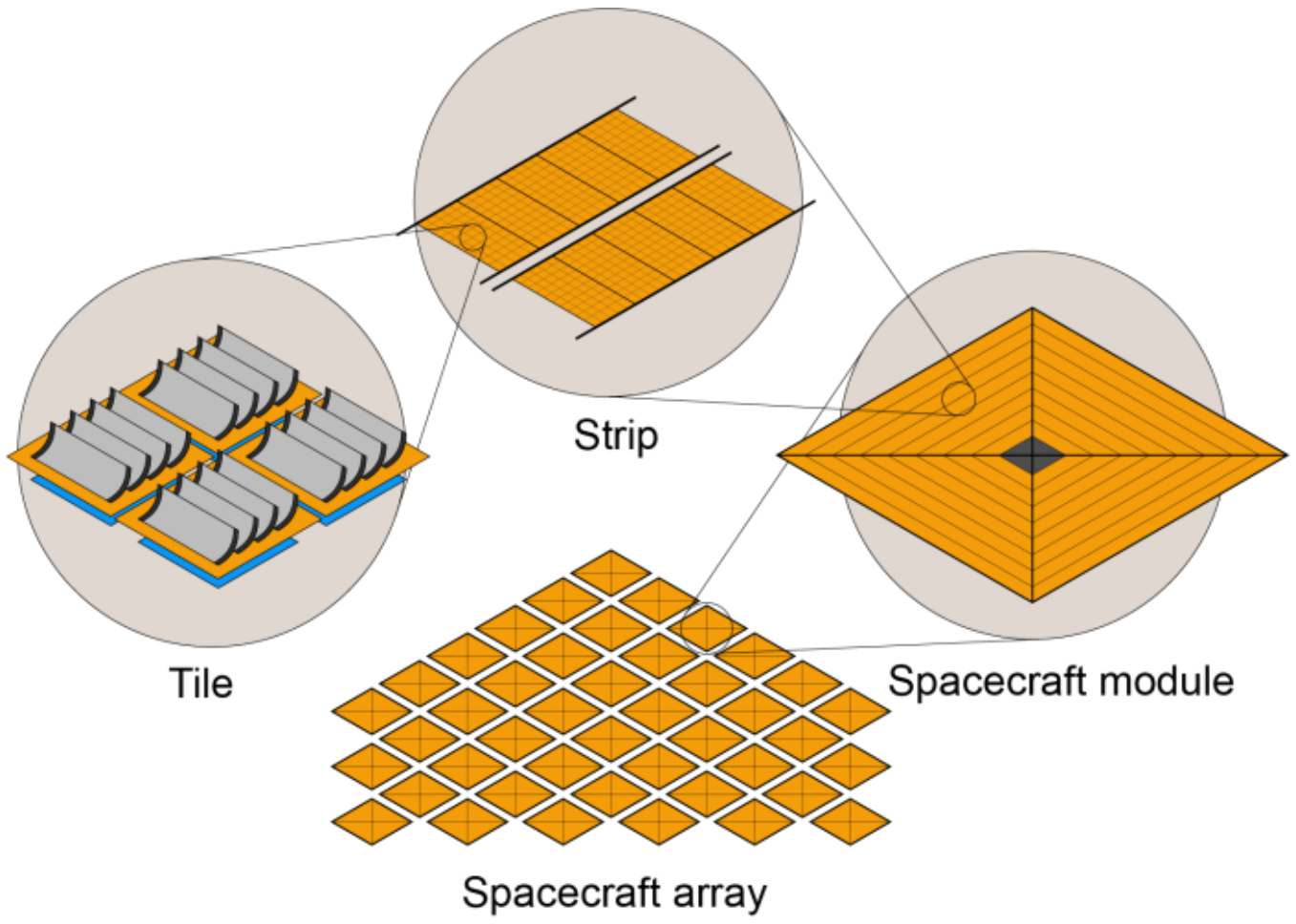

Figure 2: Conceptual drawing showing the scalability of the Caltech approach. The tile is a basic unit of functionality which is scalable by replication to larger assemblies.

## III. PHOTOVOLTAICS

### A. Specific Power and Concentration

One of the most important metrics for space-based solar cells is the specific power (W/kg), which incorporates both the efficiency of the device and its mass. Space photovoltaics have thus far been dominated by flat plate arrays deployed on their growth wafer. Although early space solar cells were made of terrestrially commonplace Si, more recently multi-junction solar cells made of III-V semiconductor materials have been used due to their higher efficiency and radiation tolerance. With solar cell efficiencies of ~30%, the most widely used space solar cells have a cell specific power of ~180W/kg and array specific power of ~70W/kg [7]. The large difference between the cell specific power and the array specific power is due to the need to include a top layer of protective coverglass to shield the cells from the harsh particle radiation environment of space. The coverglass accounts for a significant fraction of the total mass of the photovoltaic structure.

In terrestrial photovoltaics, concentrators can be used to reduce the amount of expensive photovoltaic material by the factor of the optical concentration. Using this scheme, most of the sunlight collecting area is instead covered by relatively inexpensive mirrors or refractive optics. The same concept can be applied in space not only to reduce cost of the cells but also system mass if relatively lightweight concentrators can be used. Concentration reduces the mass contribution of the solar cell by the concentration factor and, more importantly, also reduces the contribution of the radiation shield by the same factor, because the radiation shield is only needed to protect the electronic device.

**Figure** 3 shows a first-order calculation of the areal density (g/m$^2$) of a typical photovoltaic with a lightweight concentrator as a function of concentration factor. Here, we assume the concentrator is a parabolic trough reflector made of a 10 μm polyimide support coated with 10 μm layer of aluminum for reflectivity and thermal conduction. This calculation also assumes 75 μm of ceria-doped coverglass covering the cell, 30 μm of Cu metal at the rear of the solar cell, and 10 μm of active GaAs solar cell material. Figure 3 demonstrates that even with modest concentration factors, the areal density of the tile is significantly reduced with little additional benefit for concentrations larger than 20-30x, due to the ratio of mass per area of the cells vs. reflectors. Concentration increases specific power both by reducing system mass and by improving photovoltaic device efficiency [8], making concentration concepts even more attractive relative to flat plate photovoltaic designs.

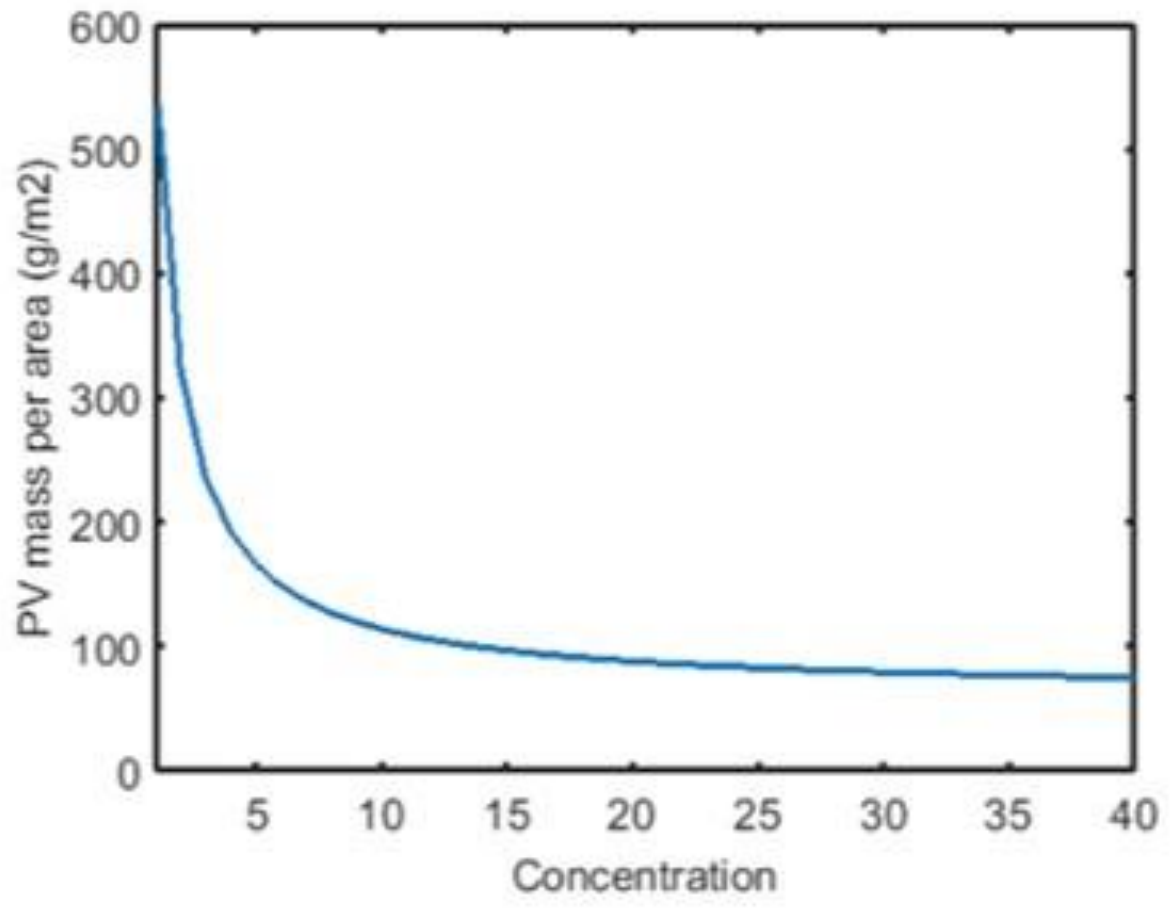

Figure 3: Areal density vs. concentration for PV + concentrator.

### B. Epitaxial Liftoff

An additional prospect for mass reduction of the photovoltaic component is thinning or removal of the semiconductor growth substrate. In high efficiency III-V based solar cells, the essential, active photovoltaic device is in fact extremely thin and lightweight -on the order of 10 µm for a typical high efficiency multi-junction solar cell. When deployed, the devices are usually still attached to their growth substrates, typically 140 µm thick Ge. The substrate, after growth, is non-essential and can easily be removed with a process known as epitaxial liftoff (ELO) [9] shown in single [10] and dual junction [11] solar cells have been prepared with this process. This not only allows up to a 15x reduction in cell mass, also renders the cell flexible and increases the overall efficiency due to increased photon recycling [12]. Additionally, the ELO process has been shown to significantly decrease the cost of III-V based photovoltaics by reuse of the growth wafer [13]. This could be a significant advantage for the large scale high efficiency photovoltaic production needed for full scale deployment of a space-based solar power system.

Figure **4**. Recently, world record

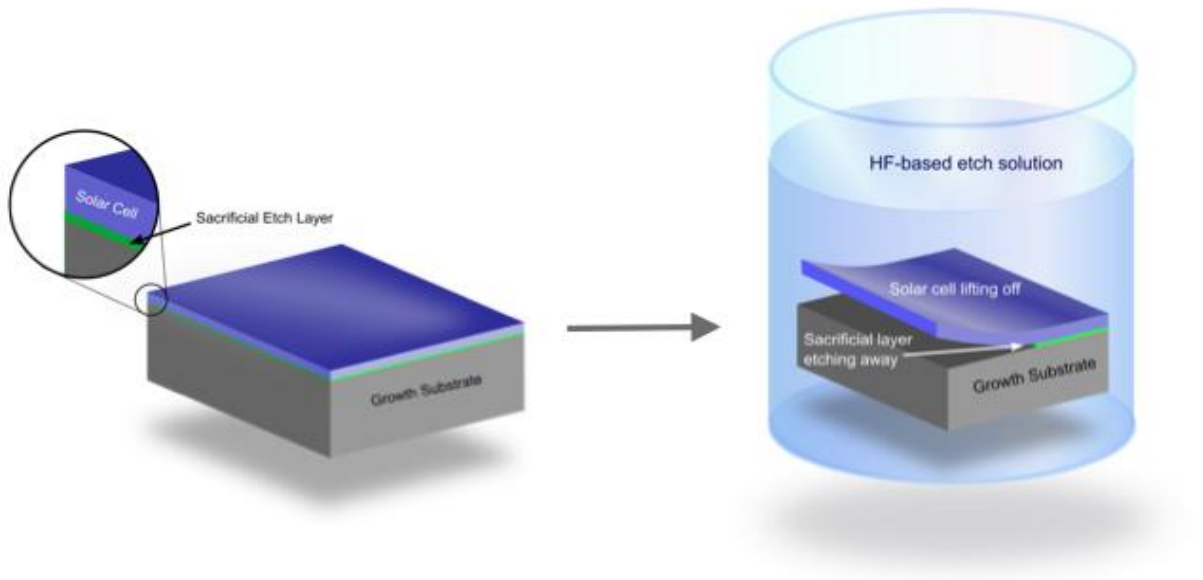

Figure 4: Schematic of the epitaxial liftoff process. The mass of the photovoltaic component can be further reduced by removing the inactive growth substrate from the device.

### C. *Concentrator Optical Design*

Our selected concentrator design is shown in Figure 5**Error! Reference source not found.**. The concentrating element is a one-dimensional half parabola reflector that focuses incident sunlight onto the back of the neighboring reflector, where the photovoltaics are located, a variant on the design of the SLATS concentrator [14]. In this application the half parabola has a few advantages over the more common full parabolic concentrator. Most importantly, it allows for the photovoltaics to be placed on a metallic, thermally conductive heat sink that provides for efficient heat extraction. This surface also acts as the solar reflector, reducing mass and structural complexity through its multifunctional design. Since only one side of the mirror needs to reflect incoming sunlight, the rear of the reflector can be coated with a custom material with high thermal emissivity, increasing the overall heat rejection and reducing the operating temperature of the photovoltaics. Additionally, the half-parabola can be stowed flat for launch and subsequently deployed more easily than a full parabola. The concentrator vanes are simply flattened when stowed and spring back to their designed shape upon deployment.

Figure 5: Schematic of parabolic half trough concentrator.

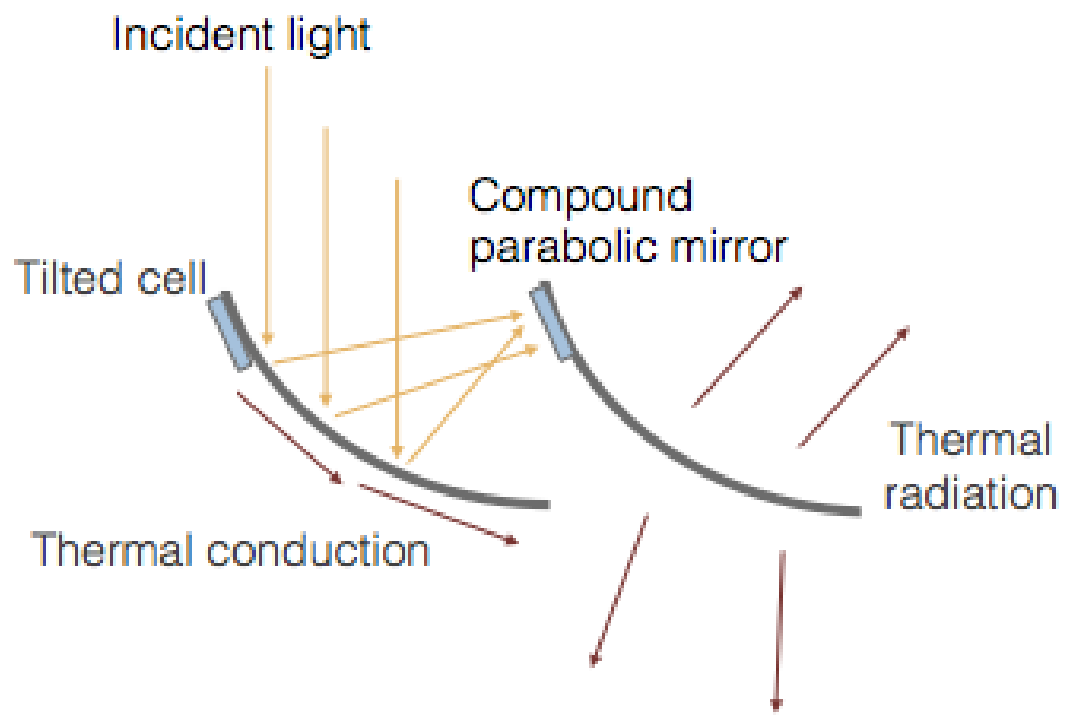

The structural portion of the reflector is made of polyimide (e.g. Kapton) with a thickness of 5 - 25µm. To increase heat rejection, a black polyimide can be used that has a high thermal emissivity (~0.88) at thicknesses of 25µm or less [15]. The sun-facing side of the reflector is coated with 1 - 10µm of Al for thermal conduction and topped with ~100 nm of Ag for efficient optical reflection. The reflectors will be held in place with the correct shape by custom carbon fiber springs, attached at various points and at the edges of the parabola.

To determine the optical performance, we performed ray-tracing simulations as a function of incident angle along both optical axes for a 20x parabolic half-trough concentrator, shown in Figure 6. In general, the angular tolerance along the concentrating axis will decrease as the concentration factor is increased for any type of optical concentrator [16], as is the case here (Figure 6-a). The optical efficiency is always less than 1 even for a perfect geometry due several factors including metal absorption and rays that miss the target due to the +- 0.26° angular spread of the sun. The angular response for the parabolic half trough concentrator design is asymmetric around normal incidence, with relatively flat response in one of the directions. For a 20x concentrator, the angular tolerance is ~1.5 - 2°, within pointing accuracy capabilities of common small satellites [17]. Along the non-concentrating axis (Figure 6-b) the angular tolerance is much higher, which could allow for tilting or slewing the satellite throughout orbit to balance RF transmission efficiency with solar collection efficiency (described in Section VI).

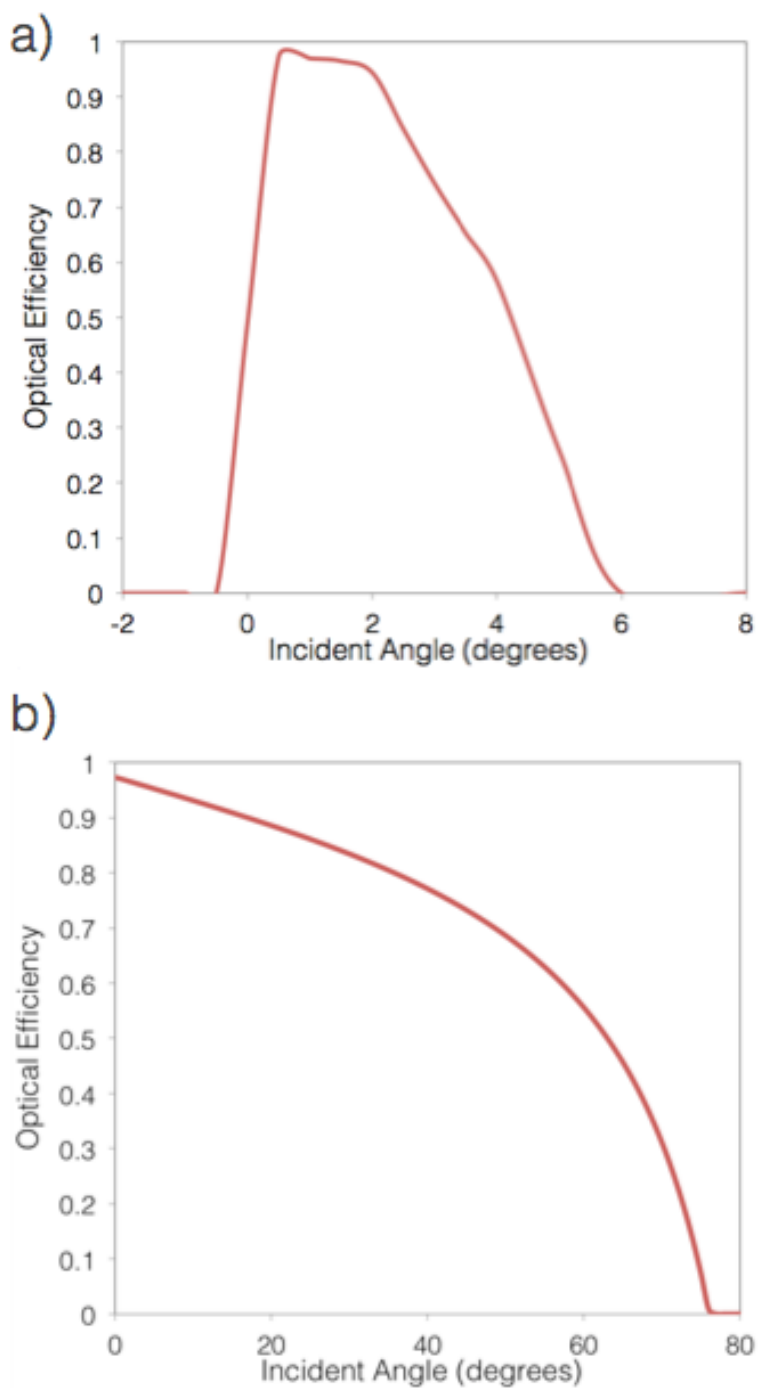

Figure 6: Calculated optical efficiency as a function of angle for 20x concentration along a) the concentrating axis, angle α, and b) along the non-concentrating axis, angle β.

## *D. Concentrator Thermal Design*

The lack of convection or external thermal conduction in space makes thermal management one of the most challenging aspects of space-based concentrator photovoltaics. Using COMSOL Multiphysics, we have calculated steady state temperature profiles for numerous variations of our design under different degrees of concentration. An effective input heat load of 650 W/m$^2$ was used, calculated from the efficiency and bandgap of state-of-the-art triple junction inverted metamorphic solar cells [18] with the assumption that we can prevent absorption of low energy photons to minimize thermal loads. We also account for expected absorption in the coverglass and in the Ag reflector. For this calculation, we also assumed a rear emissivity of 0.88 [15], front side emissivity of 0.5 (Ag coated with 4 μm of $SiO_2$ or $Al_3O_2$), and coverglass on the front surface of the photovoltaics (75um Ce doped $SiO_2$, emissivity 0.88). Figure 7 shows the temperature at the location of the photovoltaic device at 20x concentration for varying thicknesses of Al.

Typical single and dual junction solar cells prepared with epitaxial liftoff have been shown to have temperature coefficients as low as 0.1%/K [11], [19], all referenced to operation at 25 $^{o}$C. To maintain sufficient efficiency and durability, the operating temperature of the photovoltaics should be kept below 100 $^{o}$C, which is achievable for larger aluminum thicknesses as shown in Figure 7 There is clearly a compromise that must be made between mass and operating temperature, both of which will affect the specific power of the overall system.

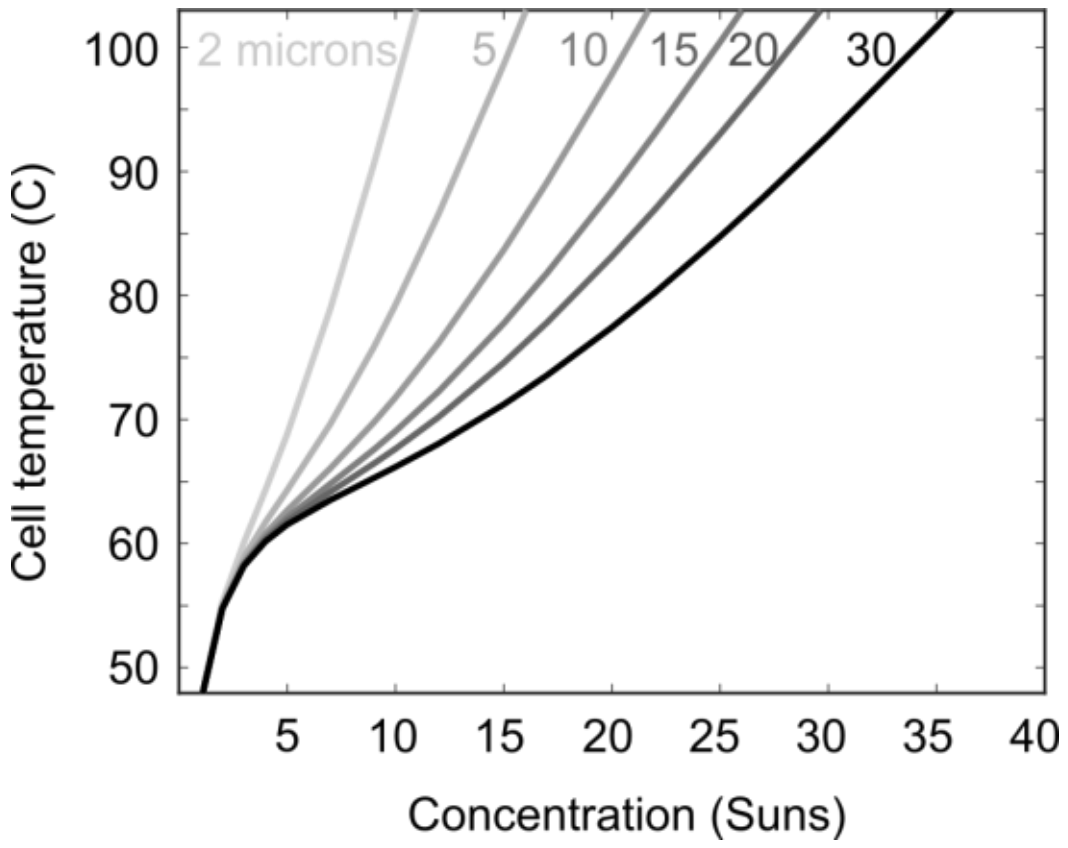

Figure 7: Cell temperature under concentration for 1 mm wide cells with a 650 W/m$^2$ heat load at one sun mounted on mirrors with 2 to 30 microns of aluminum for heat conduction. The mirror emissivity is 0.4 and the rear surface emissivity is 0.88.

### *E. Performance Estimates*

In order to estimate the overall efficiency of the photovoltaics component of our space-based power system, we performed a more detailed calculation taking into account several factors. First, we multiplied the optical efficiency obtained via ray tracing with the external quantum efficiency of a state of the art inverted metamorphic triple junction solar cell [18] and weighted it by the available photocurrent from the AM0 spectrum [20] to determine the expected current density generation of each subcell in the concentrator system (Figure 8). The current limiting junction is the top cell, with a $J_{sc}$ of 16.16 mA/cm$^2$ Using this $J_{sc}$ and values for open circuit voltage (3.04 V) and fill factor (84%) for this solar cell [Law et al, 2012], we obtain an expected overall efficiency of 29.9% for the PV system (concentrator plus solar cell) under AM0 illumination. After taking into account thermal considerations, we expect an operating efficiency exceeding 25% over a temperature range of 100-115 $^{o}$C, using typical temperature coefficients of ~0.05-0.1 absolute %/$^{o}$C.

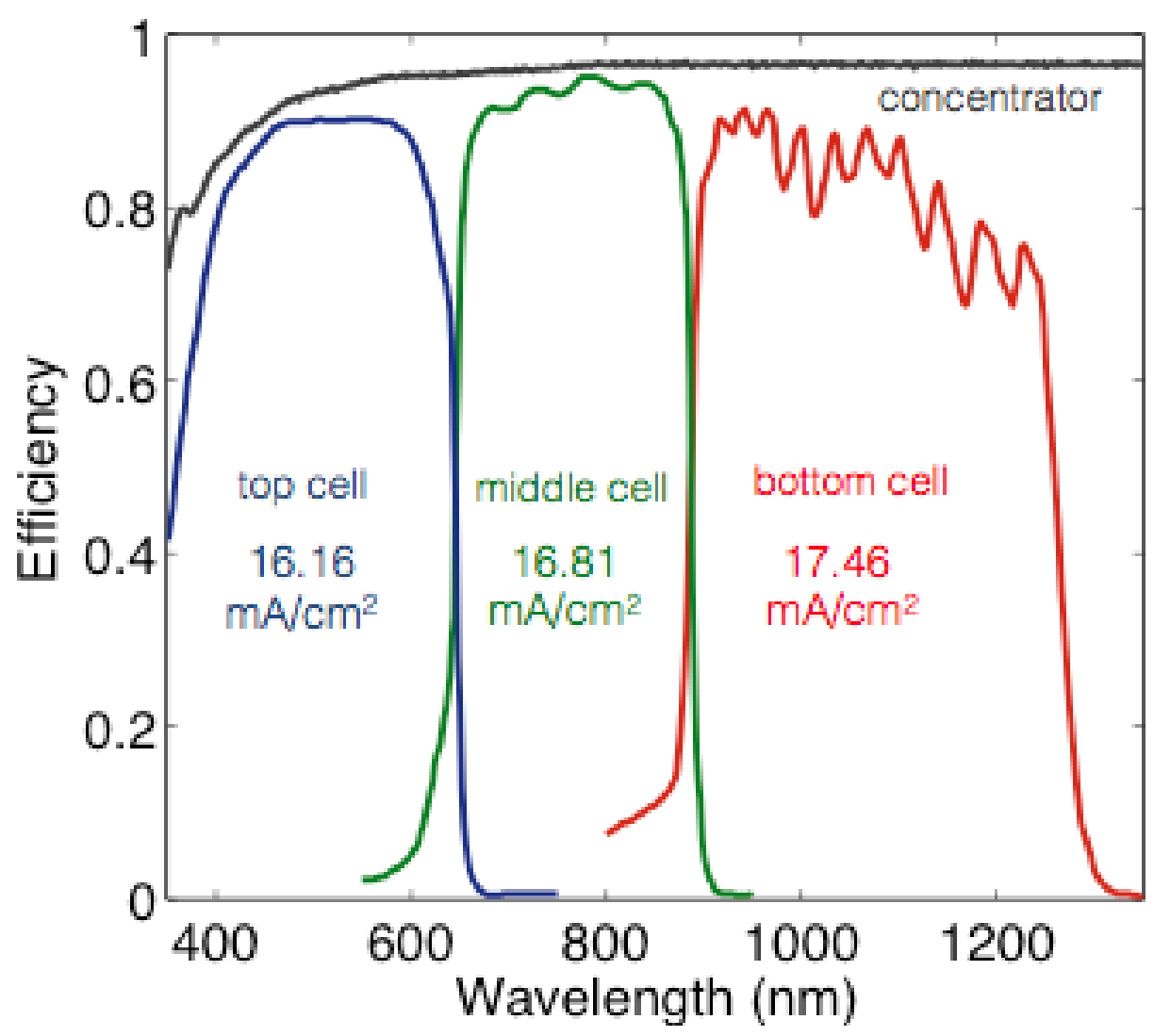

Figure 8: Optical efficiency as a function of wavelength for a 20x Ag parabolic half trough concentrator (gray) and external quantum efficiencies for each junction of a common triple junction space solar cell.

### F. *PV Summary*

In conclusion, we have designed and estimated the efficiency of a parabolic half trough concentrator and photovoltaics for use in a space-based power generating system. Optical ray tracing and numerical heat transfer calculations show that this design will allow the photovoltaics to operate at ~25-30% overall efficiency while maintaining reasonable operating temperatures under 100 $^{o}$C. The single-sided parabolic trough concentrator can be flattened to stow compactly during launch and deploy autonomously in space. The area mass density of this design is 98 g/m$^2$, which corresponds to a specific power of 3.18 kW/kg at 20 suns when a 75 µm thick glass radiation shield is included. Future design iterations have the potential to reach specific powers as high as 10kW/kg.

## IV. Microwave Power Generation and Control for Wireless Power Transfer

### A. *Introduction and Enabling Technology*

Using a large number of electronic microwave power transmitters operating with well-controlled, synchronous phase and possibly amplitude allows forming a beam that focuses to a spot in an operation analogous to that of a lens. In contrast to a lens used for focusing a beam of light, the delay of the electromagnetic radiation (or the phase for an otherwise slowly changing beam) can be electronically controlled. This independent phase and amplitude control enables a broad range of near and far field radiative patterns, generated in various directions. The electronic control of the phases enable focusing of the energy at various distances (including infinity, which would be a classical phased array transmitter). Furthermore, the ability to electronically move and steer this focus point allows for a very rapid change of the energy target.

In comparison to traditional arrays that generate the power centrally and phase shift it locally (e.g. compare discussion in [21]), generating and controlling the microwave power emission locally has several advantages, among them that no global DC or RF power routing is required and small local power density even with a large amount of total output power produced. This decentralization in turn helps with thermal management and reduces DC and RF power losses. This non-traditional mode of operation is supported by the availability of modern integrated circuit process nodes that allow economic fabrication of hundreds of millions of highly-integrated circuits that include both the RF circuits to generate and control the microwave power locally as well as the digital processing power as predicted by Moore's Law (Figure 9 [22] ).

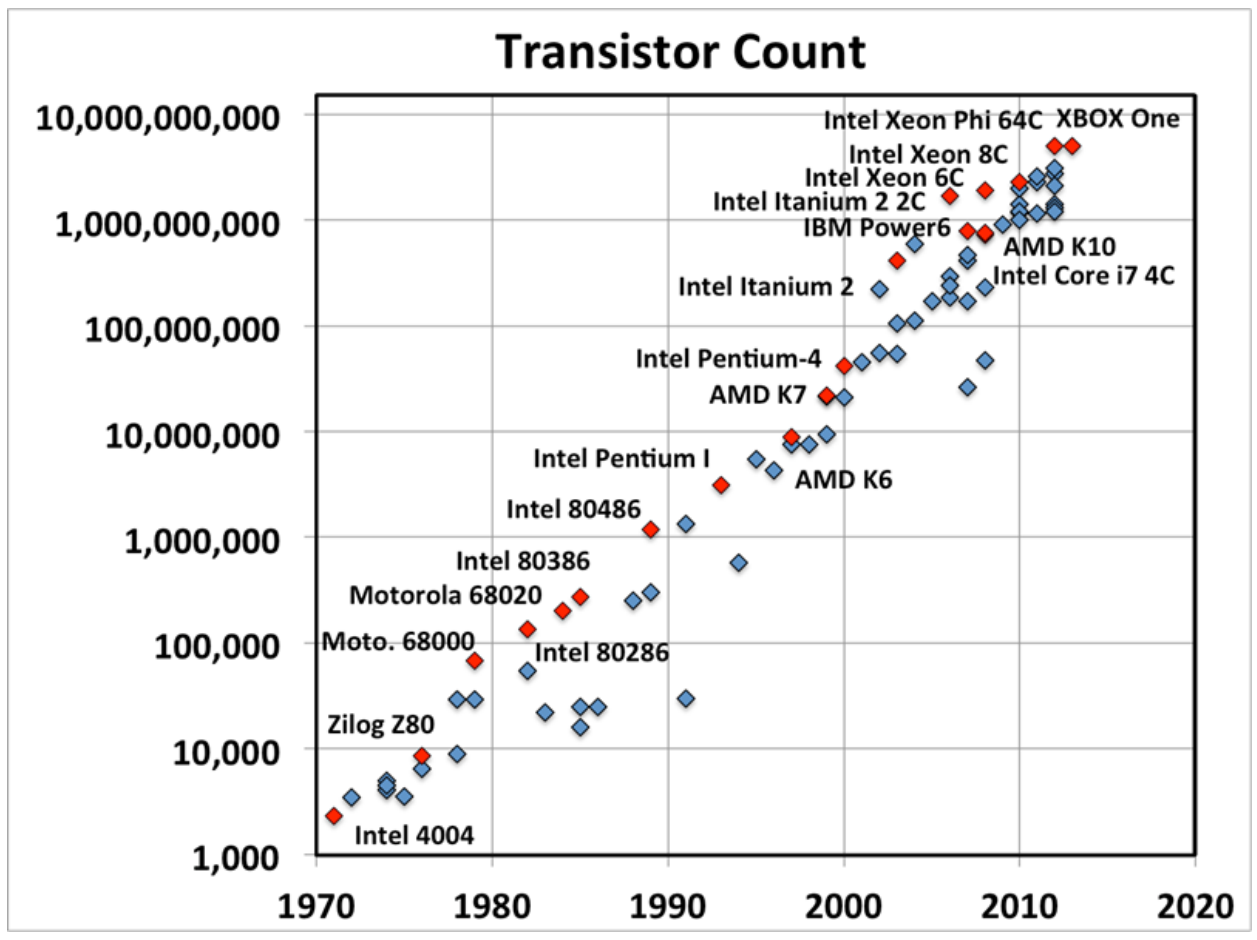

Figure 9 Transistor count of commercial CPUs by year. Figure generated by S. Bowers (University of Virginia) and A. Hajimiri, used by permission

With continuing advancements made in silicon and CMOS process technologies driven largely by the continuing need for increasing computing power, traditional approaches based on III-V technologies have, therefore, increasingly incorporated or been replaced by Silicon-based technologies (e.g. [21], [23], [24]). Traditionally, when Silicon-based technologies are incorporated, they are frequently used for back-end operations such as digital control, pre-power signal generation and receiver functionality, while transmit-receive (T/R) switches, power amplifiers (PAs) and low-noise amplifiers (LNAs) are implemented in III-V technologies in a multi-chip system solution (e.g. [24], [25]). Because of system requirements unique to wireless power transfer from space as well as continuing CMOS technology improvements, we are focusing on a fully integrated, single system-on-chip (SoC) solution to reduce cost and weight as will be detailed in later sections. Finally, as will be discussed later, performance of CMOS circuits and systems operating in an environment with ionizing radiation uniquely benefit from continuing technology scaling, compared to any other processes [26]. This also presents a strong argument for integrating all electronic functionality in one technology and SoC.

### B. *Choice of Operating Frequency*

The operating frequency of the power beaming system significantly impacts the performance, size and cost in many ways. In this section, we will discuss the most important aspects.

Everything being equal, the frequency of operation most directly affects the achievable spot size of the focused beam on Earth. Assuming an overall system efficiency $\eta_{sys}$, the diameter d of a disc-shaped system in space is:

$$\boldsymbol{d} = \sqrt{\frac{\boldsymbol{4P}}{\boldsymbol{I\pi\eta_{sys}}}} \qquad \textbf{(1)}$$

where P is the power to be available on Earth and I is the intensity of the solar radiation ($1.36 kW/m^2$). Thus, assuming η=15%, a disc with a diameter of 1.8km is required to collect the solar energy necessary to provide 0.5GW of power on Earth.

A key insight for space-based apertures of this size is that the receiver on Earth may be in the near field of the array. The far field, or *Fraunhofer Region*, is typically defined for distances of $R > 2\frac{d^2}{\lambda}$(e.g. [27]), where λ is the wavelength of the radiation. Nuances of array focusing in the near and far field are explored in [27]. At 10GHz and 36,000km distance (corresponding to the altitude for GEO above the equator), the maximum aperture size to operate in the far field would be a disc of 740m diameter, and the distance to the first diffraction minimum on the ground would be approximately given by (e.g. [28]):

$$\boldsymbol{D_1 = 1.22\frac{R\lambda}{d}} \quad (2)$$

For the parameters above, $D_1$ = 1.8km, in which contains 84% of the total radiated power. At 1GHz, a circular aperture in space of less than 2.32km would be operating in the far field. The 1 GHz, 2.32km array would exhibit a first diffraction minimum at a distance of 5.6km on the ground. In other words, the product of ground station size and space station size scales linearly with wavelength. Above 10GHz, atmospheric absorption due to water vapor around 22.2GHz can cause significant at higher precipitable water vapor levels (e.g. above 200mm) [29]. In addition, spatial and temporal control of the individual tiles becomes increasingly more difficult at higher operation frequencies due to increased electronic phasing errors for constant timing and location errors, consequently we have limited our design space to 1-10 GHz.

We note that Friis' transmission equation, frequently used to estimate power transfer efficiencies, can be inapplicable in many of these cases, because of operation close to or at near field conditions and because of the potentially large size of the receiving array on Earth.

Peak intensity at the center of the spot is related to the total power, $P_0$, the aperture $A$, and the distance, $R$, by (e.g. [28]):

$$\boldsymbol{I_p = \frac{P_0 A}{R^2\lambda^2}}. \quad (3)$$

Hence, in the example of 0.5GW of main lobe power and 10GHz as the operation frequency, the peak power intensity would be around 1.25kW/m$^2$ assuming 15% system efficiency which includes 85% efficient conversion of main lobe power on Earth to useable grid power. The density falls off to 50% and 25% of its peak value at 42% and 58% (42%·$\sqrt{2}$) of the way to the first minimum, respectively. The peak power density at the first side-lobe is <2% the peak density of the main lobe or <1W/m$^2$. Increasing the aperture in space collects proportionally more power, in addition to resulting in higher peak power concentrations, mitigated though by spreading of power compared to the theoretical Airy pattern due to near field effects.

The formation of a concentrated spot on the ground relies on the ability to accurately control the phases of the radiating elements. For random, uniformly distributed phase errors with a maximum excursion of $\pm\delta_{max}$ the efficiency compared to a perfect phase distribution is given by [30] $\eta = \frac{sin^2(\delta_{max})}{\delta_{max}^2}$ in the limit of an infinitely sized array. For normally distributed phase errors, we can calculate an equivalent $\delta_{max} = \sqrt{3}\sigma$ from the RMS error associated with incurring a minimal error at small phase offsets. This is plotted in Figure 10. Thus, even at 10 degrees RMS phase error, we still retain 97% efficiency. This is supported by independently run Matlab simulations.

Random phase errors occur due to electronic noise and incomplete knowledge and control of the physical shape of the system. Total jitter of electronic inverters/buffers is a function of total power used as well as the process technology used (with faster technologies yielding less jitter for a given amount of power consumption). Control of the physical shape is a function of the accuracy of the shape information available and the speed with which the electronics can react to changes in shape. Since changes due to vibrations are much slower than the operation speed of the electronics, measurement accuracy of shape error is the limiting factor for maintaining correct phasing. Shape determination with an RMS error of 1 mm (in the direction to the ground receiver) corresponds to 12° RMS error at 10GHz and 95.7% efficiency.

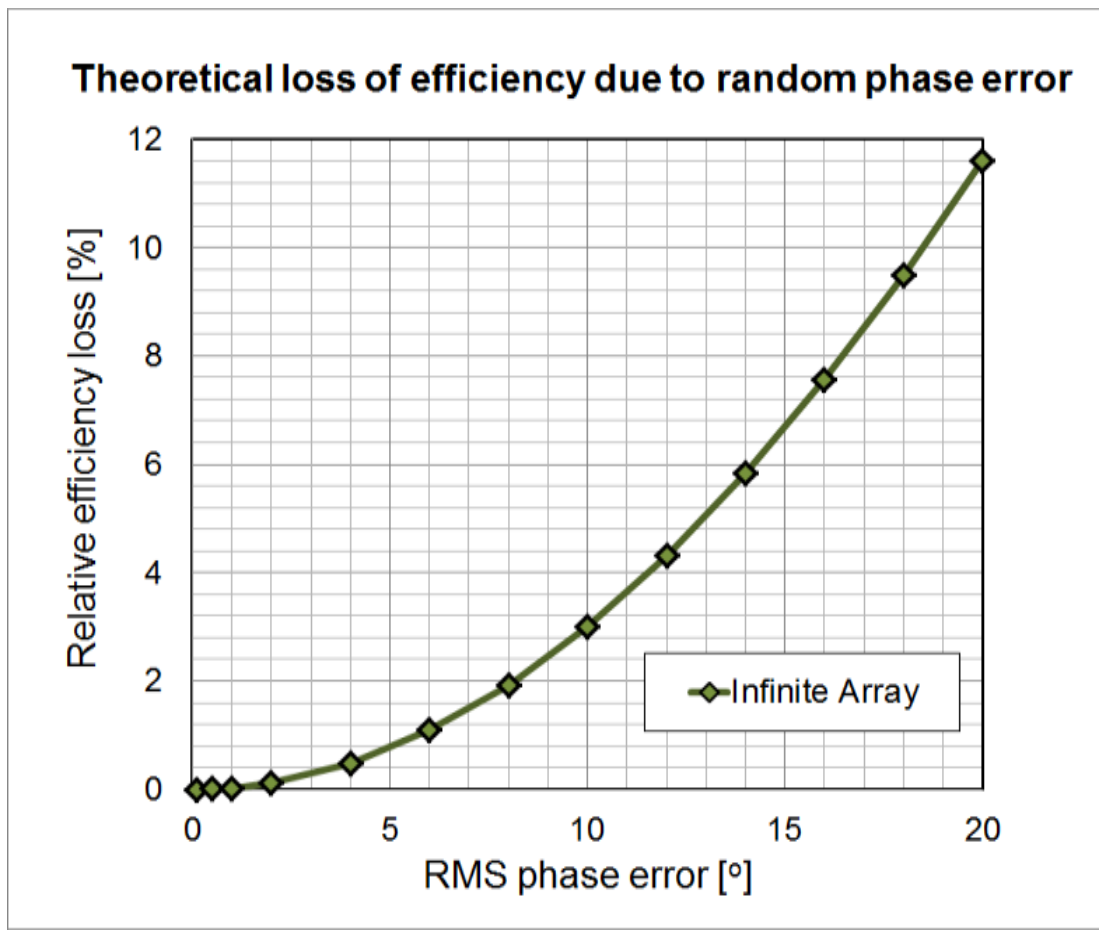

Figure 10: Theoretical loss of RF power in main lobe due to random phase errors.

One of the key consequences of our local and modular approach is that DC power generated locally will mostly be consumed and converted to RF power locally. Since insolation is constant in space, the amount of available solar power is constant per unit area of the satellite. Assuming half-wavelength antenna spacing, the absolute antenna size and spacing is determined by the choice of operation frequency, and the amount of DC power per antenna element is a strong function of the frequency of operation chosen and the efficiency of the photovoltaics (PV).

As the insolation in space is fixed, and assuming λ/2 spacing, the power per antenna is given by $\eta \cdot I \cdot \left(\frac{\lambda}{2}\right)^2$ where I is the insolation ($\sim 1.36\ kW \cdot m^{-2}$), η is the combined PV and DC-to-RF efficiency and λ is the wavelength. The nominal output power per RF amplifier is half of that amount if we generate two polarizations independently. Thus, while at 1GHz and 20% overall efficiency the nominal output power is 3W per amplifier, this amount will drop to 30mW at 10GHz, power levels that are readily obtainable even in low-voltage, advanced CMOS technology nodes. For an antenna impedance of 50Ohm, a peak voltage of $V_{pk}$=1V allows us to generate 10mW or 40mW for peak voltages of 2V (e.g. if we use cascoded transistors). We can define a power-enhancement ratio (PER) that is the ratio of the required power to the power generated over an easily realizable impedance (e.g. 50Ohm) assuming peak voltage limited operation. Using passive components, it can be shown [31] that the efficiency using n sections of passive, impedance transforming networks using inductors and capacitors is given by

$$\left(1 - \frac{\sqrt{\sqrt[n]{PER} - 1}}{Q}\right)^n \quad (4)$$

where Q is the quality factor of the inductors (assuming capacitors have much higher Q, a reasonable assumption). Thus, with a PER of 300, and a Q of 15, we would lose 40% of the RF energy in passive impedance-transformation networks for n=4, the optimal number of sections. This loss of RF energy due to higher required power-enhancement ratios counteracts the higher active efficiency at lower operating frequencies. Simulated overall efficiencies taking these effects into account for Q=20 and Q=50 are shown in Figure 11 for a representative 65nm CMOS process node.

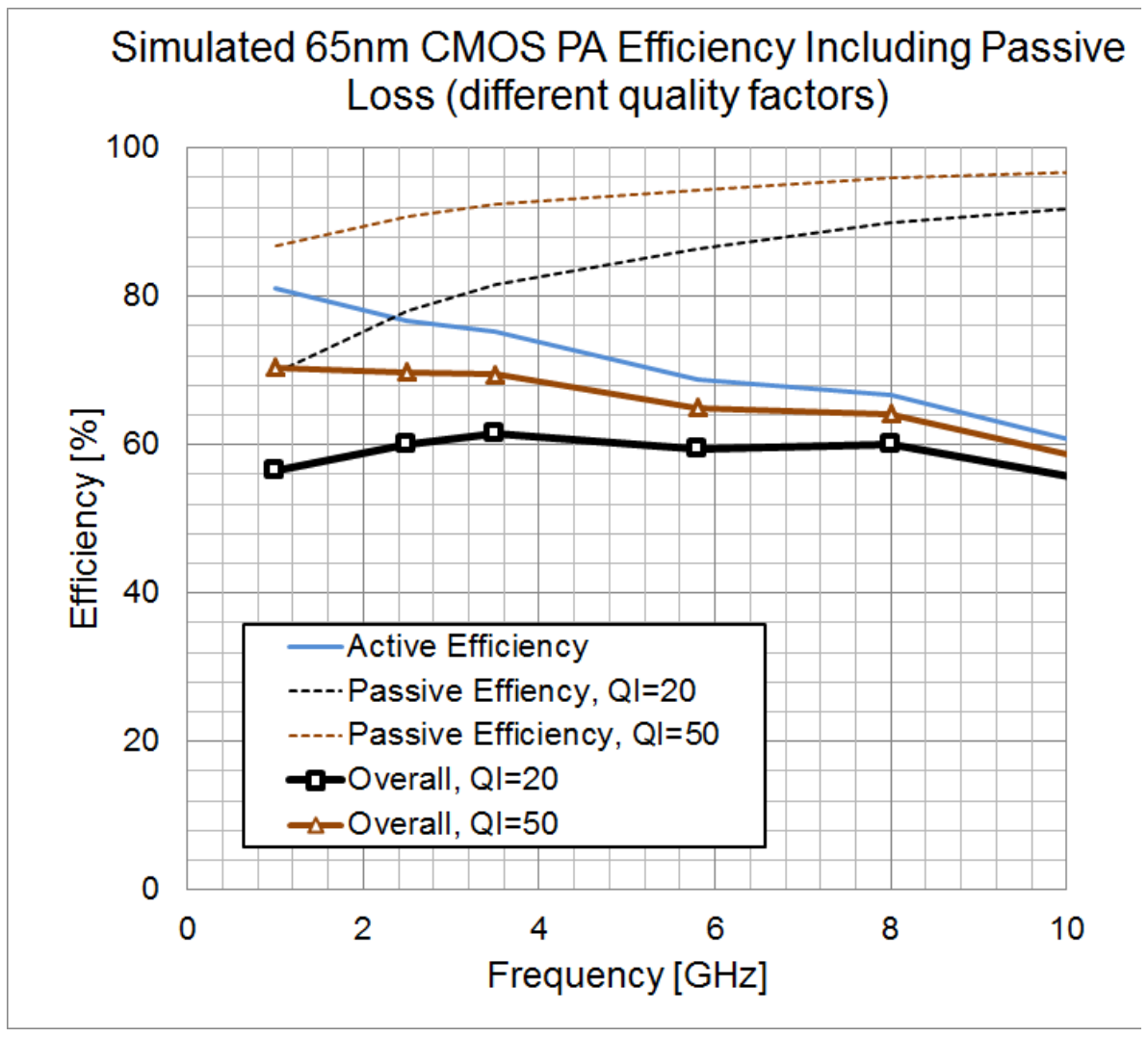

Figure 11: Simulated active only, passive only and overall PA efficiency over frequency for required output power levels and power enhancement ratios.

At lower RF frequencies and higher power levels, an option would be to use separate power amplifier device technologies such as GaAs, InP or GaN based processes that can tolerate higher peak voltages and would thus require lower PERs. However, this would create significant overhead with regards to signal routing and on-off-chip transitions, in addition to increased cost. Because power requirements even at 1GHz are within the limit of what can be achieved in CMOS technologies for power amplification [32], a single CMOS SoC promises to be the lowest-cost, lowest weight solution.

### *C. DC to RF Power Conversion in CMOS*

The bulk of DC-to-RF power conversion happens in an RF power amplifier. As a result, efficient power amplifiers are highly desirable for such task. The RF power amplifier is at the heart of the DC to RF transition, which makes its efficiency a key parameter for system performance. Because linearity of amplification is not a concern in wireless power transmission, efficient switching power amplifiers such as class E [33] [34], class F [35] or related classes of amplifiers [36] can be used. Choosing the right technology to realize the power amplifier is based on targeted efficiency, frequency of operation, RF power generated per phased array element, cost, mass, size, reliability, and radiation hardness. While traditionally integrated power amplifiers have been realized with III-V technologies such as GaAs or InP due to their larger band-gap and hence higher breakdown and output voltages, relatively recent innovations in integrated power combining topologies such as distributed active transformer (DAT) [31] [37] [32] have allowed medium power (hundreds of milliwatts to several watts) PAs with lower voltage CMOS devices with comparable power added efficiencies (PAE). MOSFETs which are mainly used as digital switches in CMOS logic circuits can be readily used as switches in switching power amplifiers. The continuous reduction of feature sizes in CMOS technology has resulted in very fast transistors which can switch over a hundred billion times a second which is the result of reduction of parasitic capacitances while keeping their switch resistance almost constant. While the reduction of transistors maximum voltage handling has resulted in lower RF power generation per device, the distributed nature of solar power and the local power conversion in the tile concept mean that each element does not need to provide high power levels. Consequently, CMOS based power amplifiers [38] can be easily integrated along with the timing generation circuitry and controlling logic in a single chip, minimizing cost and complexity of the design while increasing the robustness of RF power generation circuitry.

### D. Timing and Phase Control

As detailed in section A, the array must maintain phase coherence of the transmitters within approximately 10$^{o}$ RMS at the microwave frequency across the array to generate a phase-coherent spot on the ground. However, this requirement must be met only within the array and does not apply to full-system position errors relative to the ground station, as any systematic error in the mean distance to the space system appears as a phase offset common to all transmitters in space. An error in the other two position coordinates (elevation and azimuth) can shift the spot on the ground equal to said error, but errors on the order of meters (as achieved by GPS, for example) are insignificant compared to spot size of the microwave beam on Earth.

From this observation and noting that computational power can be made available locally on the tile level (e.g. via an integrated microcontroller or processor), a hierarchical reference distribution and phase correction scheme offers itself as a solution, shown in schematic in . In this scheme, the position of the modules relative to one another or central modules can be determined by triangulation, through the use of wirelessly transmitted reference signals (shown as module-to-module communication) and absolute clocks (comparable to the operation of global-positioning system, but for shorter distances and hence much improved accuracy). Within a module, a reference signal (red arrow) is distributed to each tile between nearest neighbors, while the motion of the tile, e.g. due to rotation or vibrations is tracked and the information is broadcast (blue arrows) from the module center to the tiles. Finally, general communication between each tile and the module control unit (center) can also be locally routed from tile to tile (green arrows). With locally available processing power on each tile, continuous correction to the arriving reference phase can be computed, predicted and applied on the tile level. The amount of data that needs to be broadcast across the module is relatively limited due to the limited number of degrees of freedom as well as the limited number of limited number of important vibrational modes.

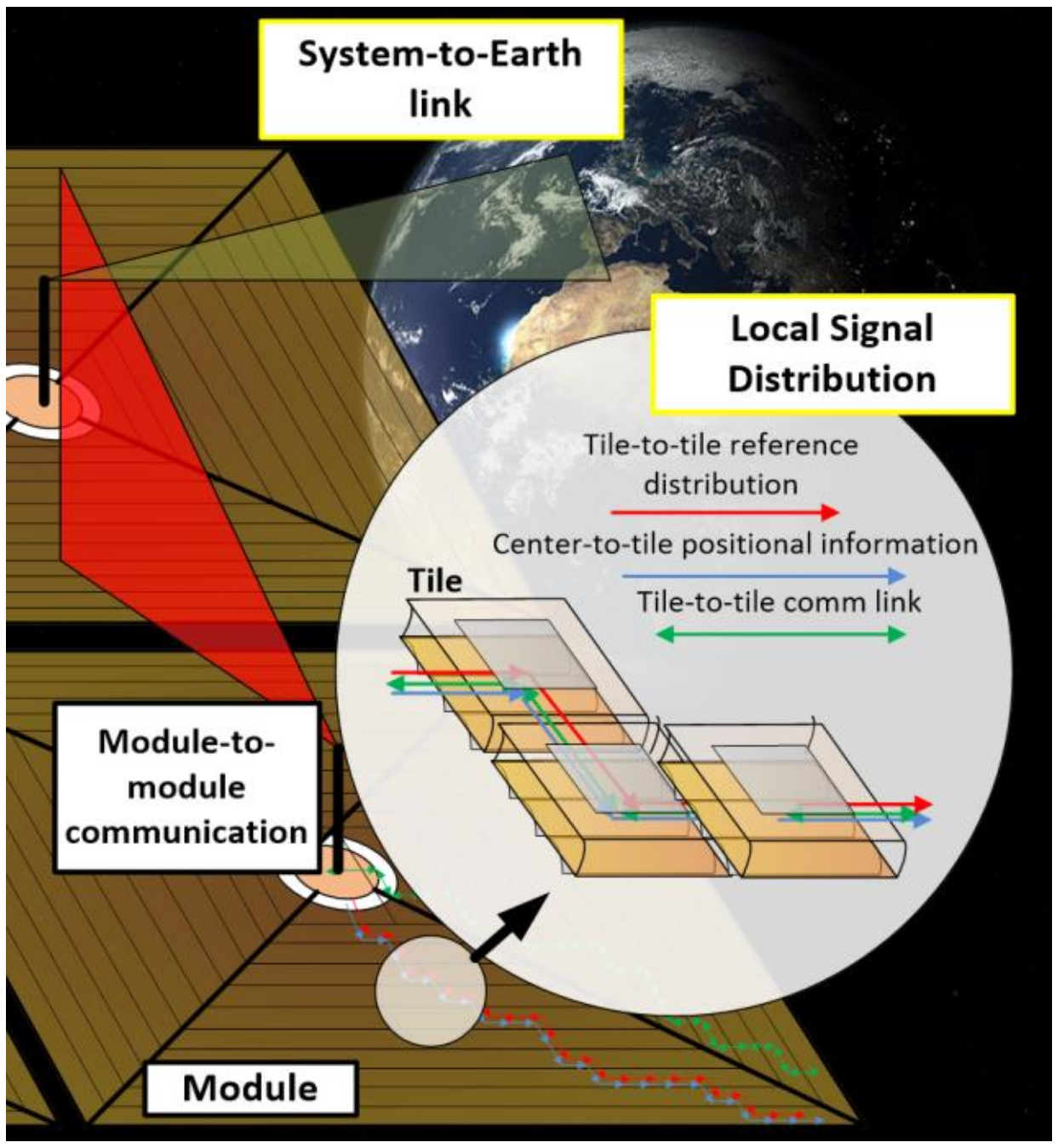

Figure 12: Hierarchical signal and reference distribution.

Because the reference phase itself needs to be distributed from a central location in the module to a million tiles or more, some form of intermediate buffering is necessary. With a module layout utilizing a tile-to-tile distribution, and a module width of W, the reference signal will be buffered on average $N = {}^{W}/_{\lambda}$ times, and the average jitter is given by:

$$J = J_0\sqrt{\frac{(N+1)(2N+1)}{2N}} \approx J_0\sqrt{N} = J_0\sqrt{\frac{W}{\lambda}}, \quad (5)$$

where $J_0$ is the jitter of an individual amplifier. If we allow the jitter contribution of the reference distribution chain to be half of the total allowable contribution (which means the distribution amplifier can contribute 12-0.52=86% of total allowable jitter), the average jitter would be 1.5ps for an individual reference buffer jitter of 47fs within the reference signal bandwidth of interest (i.e. approximately the loop bandwidth of the on-chip phase-locked loop) and a module width of 30m, which is obtainable even for reasonable power consumption. This can be further improved by schemes that use fewer distribution steps, i.e. employing a more hierarchically organized distribution approach.

Shown in Figure 13 is a simplified block diagram of the functionality within each tile of the integrated circuit system. Assuming a PV efficiency of 30%, the available DC power per antenna is 90mW at 10GHz. Using the figure-of-merit (FOM) calculations in [39] as a benchmark for a FOM of -235dBc/Hz, 1mW of power consumption results in RMS jitter 1.8ps RMS, an acceptable number at 10GHz without undue power overhead. Microcontrollers running on several mW of power consumption with CoreMark™ [40] scores exceeding 100 are commercially available [e.g. [41]] and would provide more than sufficient computing power for operational control.

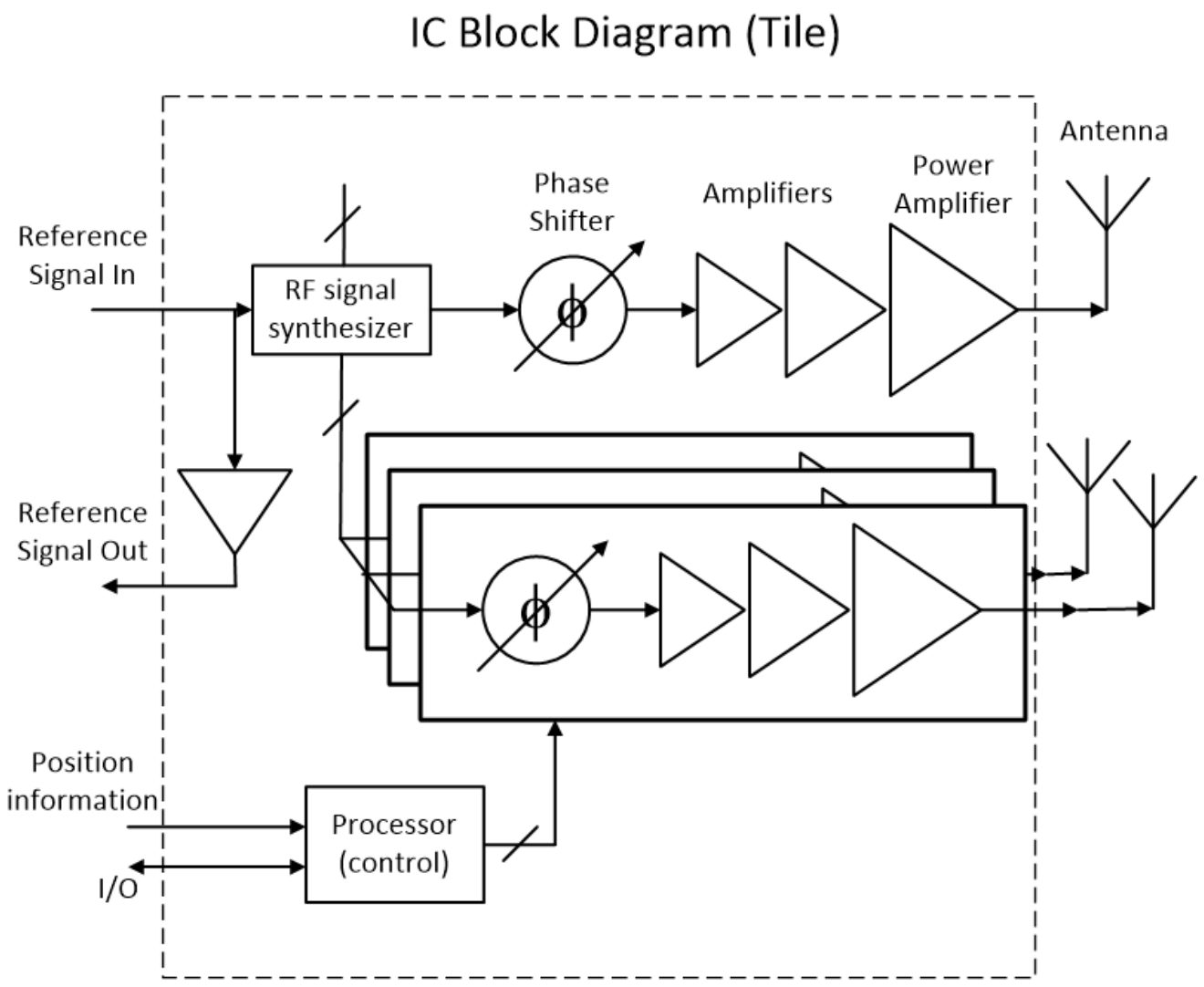

Figure 13: Simplified block diagram of RF integrated tile electronics. The reference signal is used to locally generate an RF signal and can be locally buffered for redistribution.

### E. *Antenna Design*

During the operation of the system, electromagnetic microwave power should be radiated in a beam in only one hemisphere to avoid excessive microwave power loss. While this can be achieved using a 3-dimensional arrangement that actively or passively controls radiation direction, our approach uses patch antennas combined with a reflecting ground plane to simplify stowage and deployment.

To save mass, the vacuum of space is chosen as the dielectric medium between the ground plane and the patch. This leads to a geometry shown in Figure 14.

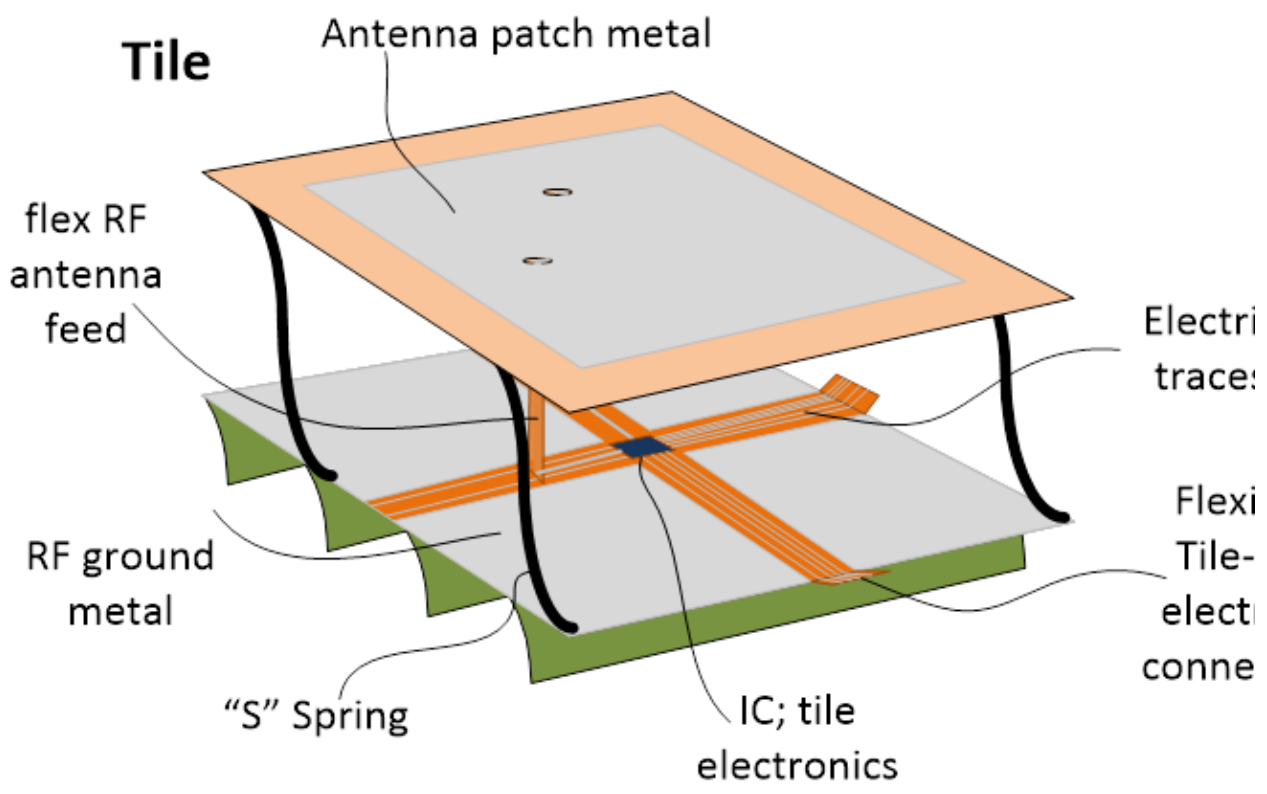

Figure 14: Tile structure (not shown to scale) showing patch antenna metal and ground layer, location of integrated circuit and the concentrator photovoltaics.

Providing two feeds to the patch antennas allows the generation of both horizontally and vertically polarized fields. Circular polarization is also achievable because the phase between polarizations can be controlled, assuming that no amplitude control is applied. Classical design formulas (e.g. [27]) indicate a patch width of close to half a wavelength, similar to the spacing preferable for the antennas in the arrays themselves (to avoid additional grating lobes). The element antennas can be sized somewhat smaller to reduce mass, however, because the presence of the adjacent array antennas naturally tunes the array to the half wave-length spacing, and because grating side-lobes will be small for the relatively shallow steering angles used in the system.

The antenna efficiency is predicted to be in excess of 95% based on simulations of losses due to surface imperfections in the metal film. With 5 µm thick Al metal patches suspended on an 8 µm Kapton HN film, the antenna plane has an areal mass density of 20 g/m$^2$ and the ground plane with 5 µm of Al at full coverage on an 8 µm Kapton film has an areal density of 25 g/m$^2$.

### *F. Thermal Management*

Conversion losses in the DC-to-RF conversion process result in waste heat that must be dissipated to the environment. For this design, with 20% efficient photovoltaics and 60% efficient DC-to RF conversion, the waste heat corresponds to roughly 110 W/m$^2$ or 1.1 W per tile. This heat must conducted away from the IC and radiated to space while maintaining the IC at an acceptable temperature. While the IC chip will be mounted on the metallized ground plane, the 5 µm Al thickness is not sufficient to conduct heat from the chip and maintain the temperature below 100 °C, thus we incorporate additional Al thickness for thermal management. We have optimized the additional metal to provide the needed thermal conductance in a mass-efficient way.

Three metallization profile scenarios were considered: the first profile has quadratically increasing metallization thickness from a thickness of zero at the tile edge ("zero edge thickness"). The second profile keeps the total metal volume constant underneath any constant width annulus around the center ("equal volume"). For the third profile, a Nelder-Mead optimization is run to select the heights of five different plateaus around the center. The thickness z in mm versus the distance x from the center in mm is shown in 7 for all three profiles. In all scenarios, the total volume (mass) of metal used is held constant at the volume occupied by a constant thickness metallization of 5um (50mm$^3$), 10um or 2.5um.

In order to decouple potential issues with RF-IC heat dissipation from other parts of the system a simplified model was examined, in which a source of heat, modeling the IC, heats a thin aluminum sheet on a thin, grey body material, modeling the polyimide structural material. In order to decouple potential issues with RF-IC heat dissipation from other parts of the system a simplified model was examined in which the IC is modeled as a heat source. A variable-thickness aluminum sheet and a thin Kapton layer to represent the heat-conducting circuit board materials. A background temperature of 30K is assumed. Radiation occurs into one hemisphere effectively assuming that the PV side is blocked from effectively radiating heat. The problem setup constrains the tile dimensions and optimizes the aluminum thickness profile to achieve maximum heat dissipation for a given tile mass. As a result, the aluminum sheet is thick at its center and thins monotonically towards the tile's boundary.

The simulation setup is shown in Figure 15. For simplicity, radial symmetry was assumed in all simulations.

Figure 16 shows the simulated IC temperature for these three profile assumptions for 2W and 0.5W per IC, equivalent to operation at 1GHz and 2GHz (accounting for the reduced tile area). As expected, temperature decreases as the sources of heat are more evenly distributed over the available area (corresponding to more sources producing less heat, equivalent to operation at higher RF frequencies). The Nelder-Mead optimized profile is the most efficient metallization profile of the three profiles investigated.

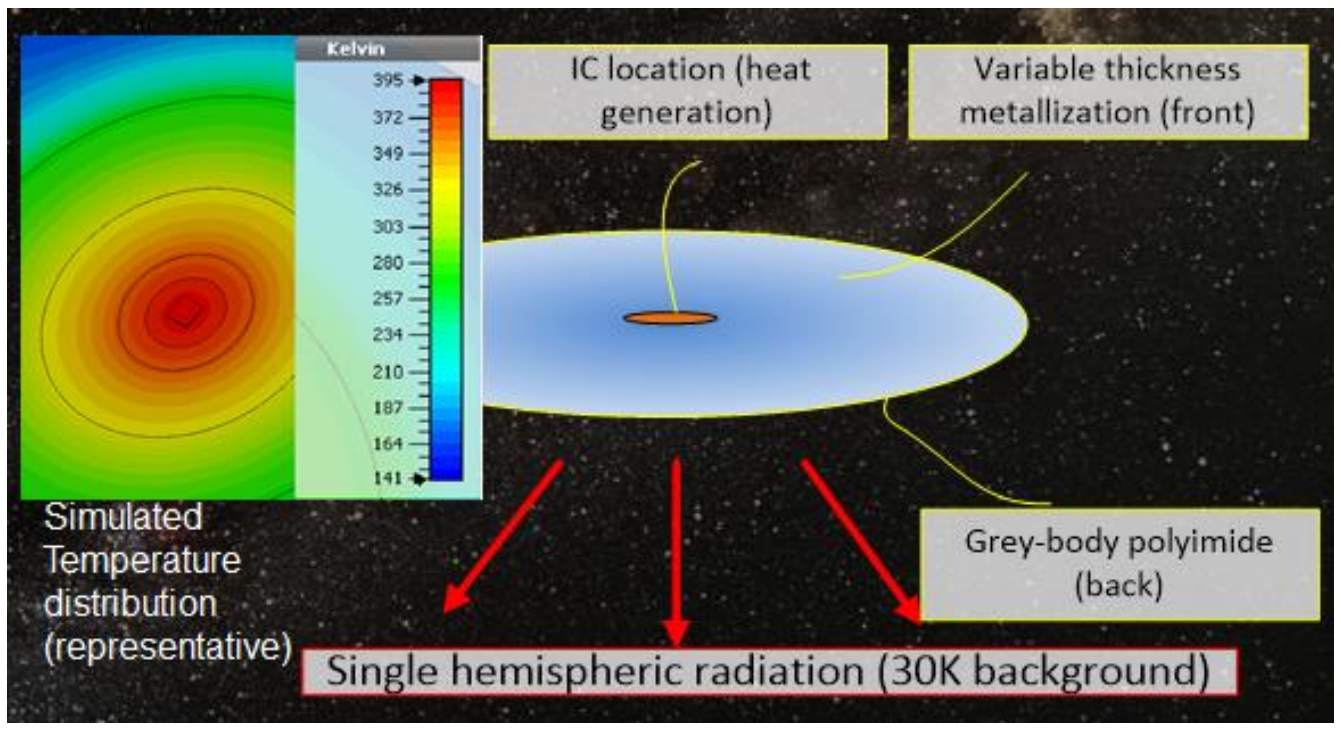

Figure 15: Simulation setup for simulating the temperature rise of the integrated circuit in the tile.

Shown in Figure 17 is the effect of doubling and halving the total amount of metal used, again as a function of amount of heat per source (IC). The simulations make the "equal volume" assumption to reduce simulation time. Compared to constant thickness metallizations, the IC temperature is reduced by hundreds of degrees Kelvins in all scenarios and acceptable under various metallization and power dissipation scenarios. The additional 50mm$^3$ of Al per tile adds 13.5 g/m$^2$ to the ground plane areal mass density, bringing it to 38.5 g/m$^2$. Combined with the antenna plane and the mass of the ICs, the RF subsystem has an areal mass density of 60.5 g/m$^2$.

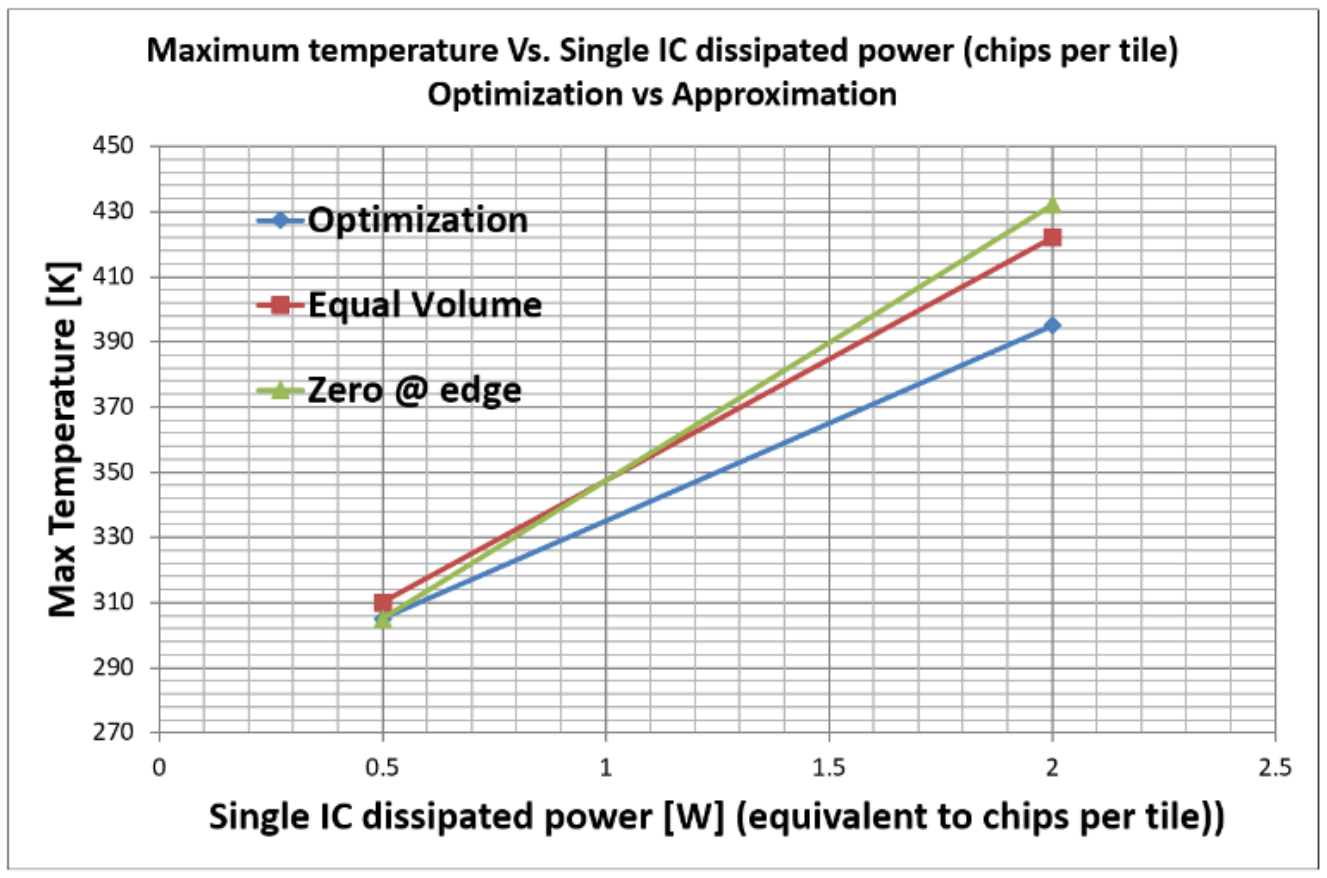

Figure 16: Simulated IC temperature for three metallization profiles using a total Al heat

conducting mass of 5um constant thickness equivalent.

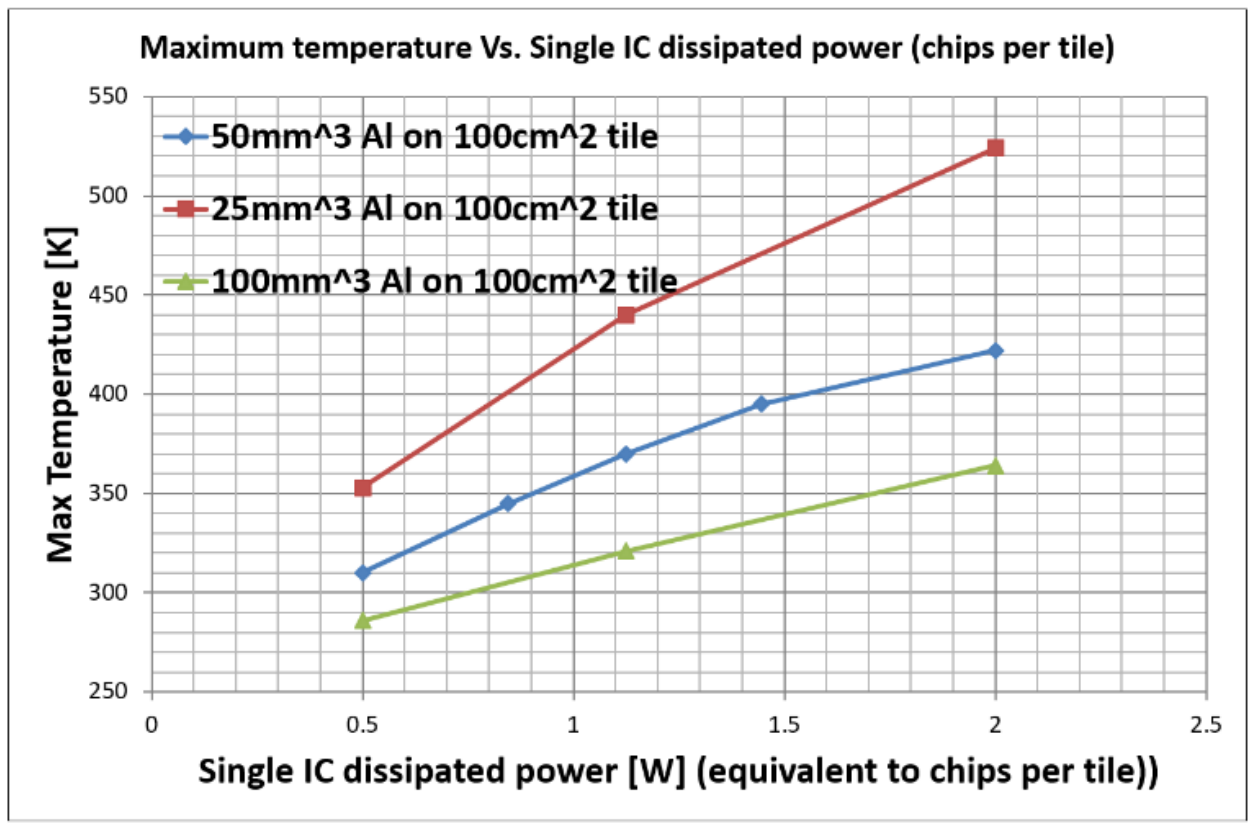

Figure 17: Simulated IC temperatures using half and twice the amount of metal for heat conduction.

### G. *Radiation Effects*

Tolerating the harsh radiation environment in space and ensuring acceptable system lifetime poses additional challenges compared to ground-based designs. For system-lifetimes of 10yrs, total ionization doses measured in Mrad may have to be absorbed and tolerated by the electronics even if noticeable radiation shielding is used (compare discussion in section V-D.). In addition, designs need to tolerate and correct for single-event effects such as corruption of random-access memory data due to, for example, a cosmic-ray strike. An additional challenge to designing for radiation tolerant systems using integrated circuit processes is the observed variability in achieved tolerance within the same lot or even on the same wafer [42].

Compared to other technologies, modern small-feature size CMOS technology nodes have shown to be more capable of handling stress due to ionizing radiation, as well as to have more promise to benefit from technology scaling compared to bipolar and III-V processes [26]. Measurements of aging in the presence of radiation have shown that the change in threshold voltage gets smaller as the feature size of the technology reduces [43]. This can be attributed to the fact that advanced node technologies have much lower gate oxide thickness which reduces the probability of a radiation generated electron being trapped in the oxide and changing the threshold voltage [26]. Also the oxide thickness in many advanced technologies is so thin that the electrons can tunnel through and produce a leakage current. While this effect is an undesired phenomena in digital circuit design as it introduces static power dissipation, tunneling of electrons increases the tolerance of the MOSFETs to radiation as the electrons inserted in the gate oxide do not stay trapped and the threshold voltage recovers after the radiation event. While thick-oxide interfaces (e.g. field-oxide or shallow-trench isolation) are more vulnerable to ionizing radiation than gate oxides for deep-submicron CMOS processes, these effects can be greatly mitigated by choosing custom layout techniques. Enclosed transistors and closed gate structures can prevent increased leakage currents and hence power consumption over time in radiative environments [44] [45]. In addition, periodic thermal annealing can be adopted to recover radiation-induced performance degradation [43] [45]. Using these techniques, analog integrated circuits capable of withstanding tens of Mrad TID can be implemented in commercial CMOS processes [45] [46].

Finally, single-event effects in CMOS circuits [47], while not causing long-term damage, affect the operation of digital circuits by corrupting digital information. As with the analog portion of the circuits, design and layout techniques for memory cells [48] as well as integrating redundancy and error-correction methods into the ASIC design flow (e.g. [49]) can mitigate these effects and result in greatly improved robustness in operation. Due to the large number of circuits in the space-based system, the (temporary) failure of any individual tile has an insignificant effect on overall performance, and periodic global status monitoring can detect local, non-destructive faults and correct them (e.g. via a reboot).

## V. Structures, Packaging and Deployment

We have developed a structural concept for a spacecraft module that coils tightly into a cylinder and deploys into a large square structure. It uses ultralightweight components connected by simple joints that can be mass produced.

The key structural design requirement arises from the narrow angular tolerance of the half-trough concentrators. Based on the results presented in Section III-C, it was required that the planarity of the whole surface should conservatively have a slope no greater than 1° in any direction.

The microwave phased array described in Section IV-D can compensate, through phase corrections of the antennas in each tile, for any non-planarity of the structure. An in-space metrology system based on sun sensors distributed over the structure has been proposed. These sensors measure the relative angles from the sun and the shape of the structure can then be reconstructed with an accuracy on the order of a millimeter. [50].

Therefore, the structural design considers only the effect of angular deviations from the nominal planar configuration of the concentrators.

### A. *Module Packaging Concept*

The initial inspiration for the packaging concept came from packaging techniques that have achieved tight packaging of continuous solar sail films with circular [Guest and Pellegrino] or square [51] shape. Even tighter packaging, that also avoids permanent deformation of the film material, can be achieved by dividing the sail film into straight strips of equal width, connected by slipping folds [52]. Allowing relative sliding between adjacent strips accommodates the increasing radius of the coil due to the thickness of the film. A fully connected edge of the film can be achieved if the slip along the folds is a symmetric function with zero-value on both ends of the slipping fold. This can be achieved by inverting the curvature of the coiled strips at the center, see [52] for details.

Based on these considerations, the overall structural architecture that we have selected is a square divided into concentric strips of equal width. Figure 18 shows a conceptual illustration of the module fold pattern for the case of 5 strips. It consists of concentric, equally spaced folds, alternating between mountain and valley folds. Additional folds run along the diagonals of the squares.

Folding along these lines produces a star-like shape with four *arms*, as shown in Figure 18c. Wrapping these arms results in a compact packaged cylinder (Figure 18e). There are five voids in the packaged form, one in the center and one at the root of each wrapped arm, but if the strips are long the volume of these voids is small in comparison with the total volume of the coiled structure. The slipping folds allow for adjacent strips to slide past each other, accounting for the different radii of the strips in the wrapped configuration.

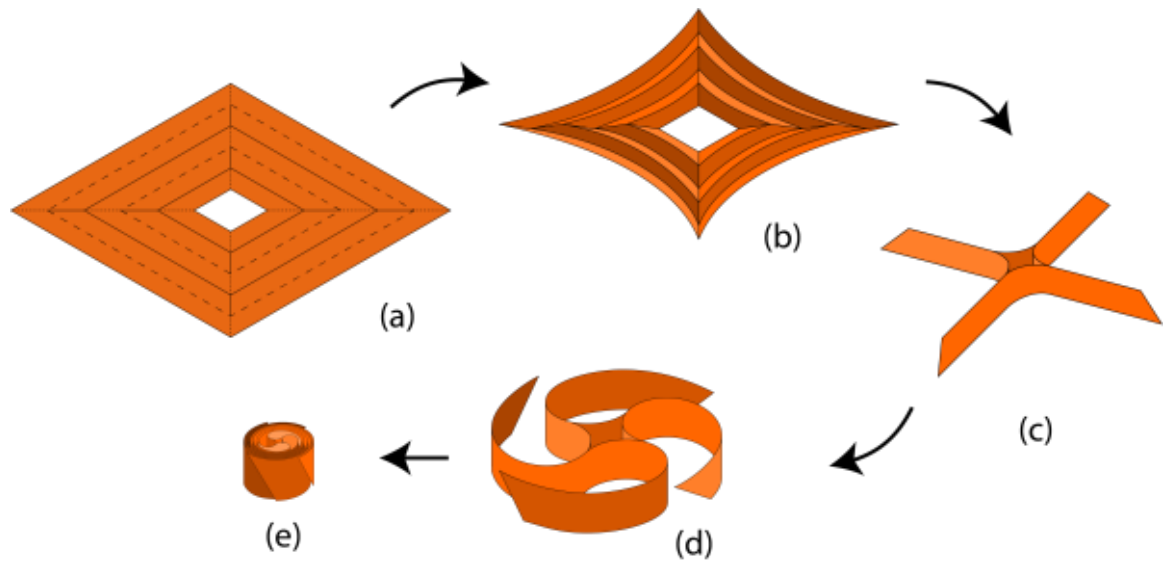

Figure 18: Spacecraft packaging concept. For clarity, only the outermost strips are shown in (d) and (e), and the scaling is increased.

## B. Loads

The external loads that cause bending of the deployed spacecraft include solar radiation pressure, inertia-related forces related to attitude control maneuvers, and gravity gradient effects.

The solar radiation pressure, SRP, at 1 AU from the sun, assuming a perfectly reflective surface with a solar incidence angle alpha relative to the surface normal, is given by [McInnes, 1999]:

$$\boldsymbol{SRP} = \frac{2AM0}{c}\cos^2\alpha \qquad (6)$$

where $AM0 = 1366$ Wm$^{-2}$ is the solar energy flux at 1 AU, $c = 3.00 \times 10^8$ ms$^{-1}$ is the speed of light, and the factor 2 assumes a perfect reflection. For $\alpha = 0°$, $SRP = 9.1 \times 10^{-6}$ Pa

The inertial forces due to the rotational accelerations can be estimated by assuming a minimum-time acceleration profile to slew the satellite about an arbitrary axis through an angle Δθ over a given time Δt. Assuming the angular acceleration to have the value $\ddot{\theta}_{max}$ for Δt/2 and $-\ddot{\theta}_{max}$ for the remaining Δt/2, the maximum slew velocity is $\dot{\theta}_{max} = \ddot{\theta}_{max}\,\Delta t/2$ and the slew angle is related to $\ddot{\theta}_{max}$ by:

$$\Delta\theta = \frac{\ddot{\theta}_{max}\Delta t^2}{4} \qquad (7)$$

Therefore, the maximum possible normal inertia loading, at the corners of the structure, has the magnitude:

$$S = \rho\frac{2\sqrt{2}L\Delta\theta}{\Delta t^2} \qquad (8)$$

where $\rho$ is the structure areal density. For a 90° slew maneuver lasting 1 h, the maximum normal inertia loading on a spacecraft with areal density of 160 g/m$^2$ and side length $L$ = 60 m is $S = 3.3 \times 10^{-6}$ Pa.

The gravity gradient loading results from changes in the gravitational field. The gravitational acceleration at a point defined by the position vector **R** from the center of the Earth to the point has the expression

$$\boldsymbol{a} = \boldsymbol{GM}\,\mathbf{R}/\boldsymbol{R}^3 \qquad (9)$$

where $G = 6.674 \times 10^{-11}$ N m kg$^{-1}$ is the gravitational constant and $M = 5.972 \times 10^{24}$ kg is the mass of the Earth.

The component of the gravity gradient that causes bending of the spacecraft at a point defined by the local vector **r** such that $\boldsymbol{R} = \boldsymbol{R}_0 + \boldsymbol{r}$, see Figure 19is given by the linearized expression in [Ashley 1967]

$$\Delta\mathbf{a} = \frac{\mathrm{GMr}}{\mathrm{R}_0^3}\frac{3\sin 2\theta}{2} \qquad (10)$$

Multiplying by the areal density, the normal inertia loading has the expression

$$\mathbf{GG} = \rho\frac{GMr}{R_0^3}\frac{3\sin 2\theta}{2} \qquad (11)$$

which reaches its maximum when $\theta = 45°$ and $r$ is largest, at the corner of the spacecraft. For the same spacecraft considered above, in GEO ($R_0 = 42{,}164$ km) the largest value is $GG_{max} = 6.8 \times 10^{-8}$ Pa.

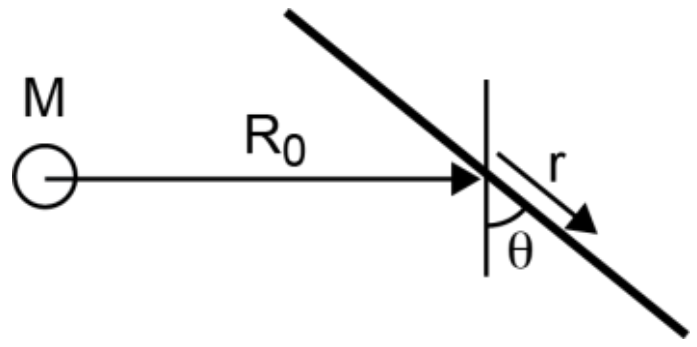

Figure 19: Geometry of spacecraft for analysis of gravity gradient loading.

Of the three loads described above, the solar radiation pressure is significantly larger and therefore will be used as the main loading case for preliminary structural design of the spacecraft.

*C. Module Structural Architecture*

We have considered three structural architectures for the spacecraft module, all compatible with a square configuration that provides minimal obscuration of sunlight and RF radiation. Overall, the structural architectures consist of strips arranged in concentric squares, as shown in Fig. 20, and connected at either end to *diagonal cords*. The diagonal cords are attached at one end to a central hub and at the other end to the tips of deployable booms. The booms are deployed from the center hub and are parallel to the diagonal cords. The performance of the concentrators in the tiles depends on the local sun angle. As shown in Figure 20, the local sun vector can be decomposed into a component within the plane of concentration, which makes an angle α with the local tile normal, and a component perpendicular to the plane of concentration, which makes an angle β with the local normal. The optical efficiency of the concentrators depends on α and to a lower extent β. In the present study the concentrators are arranged to be all parallel across the entire spacecraft. This allows the spacecraft to slew in a manner that changes the global β angle without changing the global α angle, thus minimizing the effect of such slews on concentrating efficiency.

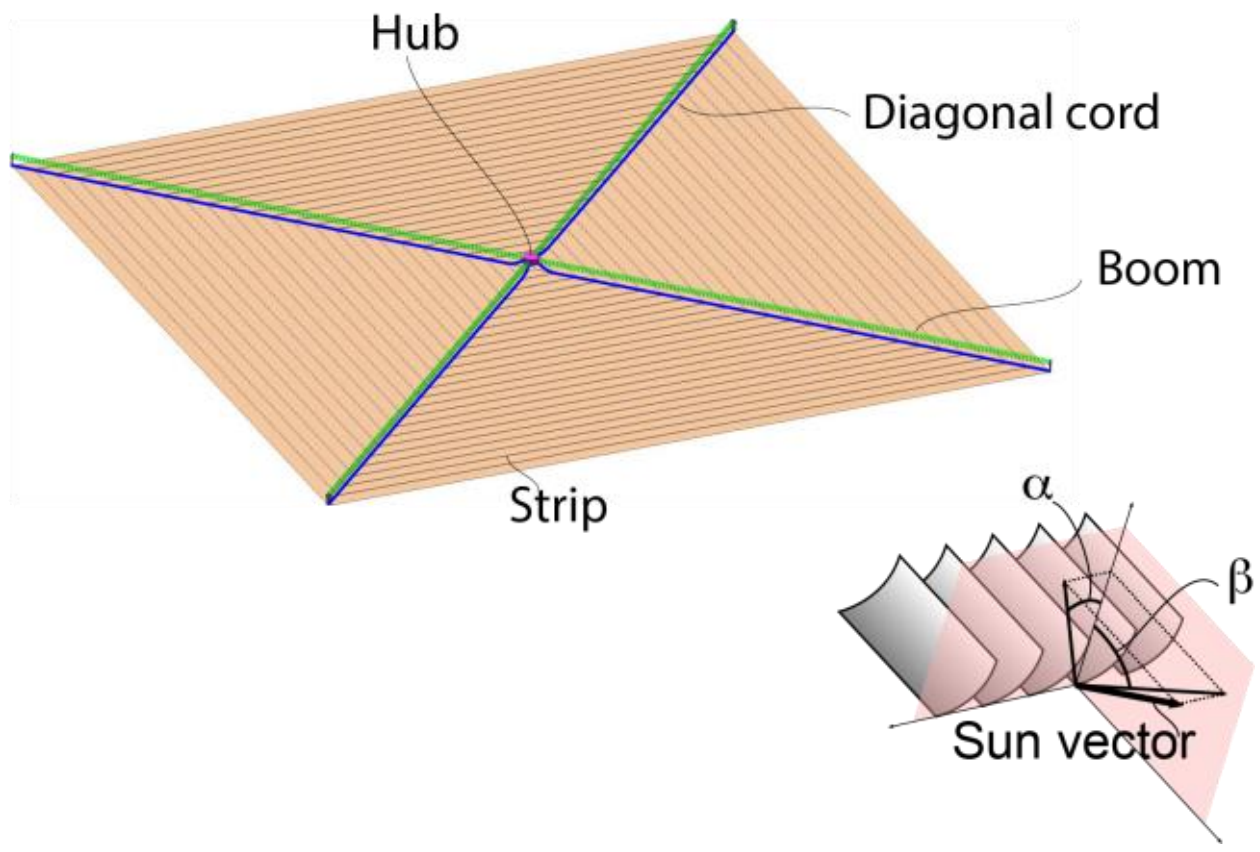

Figure 20: Overall structural architecture. The sun vector for each tile is decomposed into a component in the plane of concentration, at an angle α to the tile normal, and a component perpendicular to the plane of concentration, at an angle β.

The first architecture is based on the concept that the strips have negligibly small flexural stiffness and are prestressed to achieve the required out-of-plane stiffness. The strip prestress is 4-fold symmetric about the axis of the spacecraft. Denoting by $N_i$ the axial force in the strip, whose value is such that the slope at the ends is $\alpha$, the cords are loaded by the resultants of the pairs of cord forces, and are also loaded by in-plane and out-of-plane force components at the hub, $N_h$ and $S_h$, as shown in Figure 20.

The axial force at the outer tip of the cord is then given by:

$$\boldsymbol{N_b} = \frac{\boldsymbol{n\,SRP\,w}}{\boldsymbol{2\tan\alpha}}(\boldsymbol{2\,r_h} + \boldsymbol{n\,w}) + \boldsymbol{N_h} \tag{12}$$

where $w$ is the strip width and $r_h$ the hub radius.

The second architecture is based on the concept that the flexural stiffness of the strips provides enough stiffness that no prestress is required. Denoting by $EI$ the flexural stiffness of the strip, the slope is given by:

$$\boldsymbol{\alpha}(\boldsymbol{x}) = \frac{\boldsymbol{SRP\,w\,L_i^3}}{\boldsymbol{EI}}\left(\frac{\mathbf{4}}{\mathbf{3}}\left(\frac{\boldsymbol{x}}{\boldsymbol{L_i}}\right)^{\mathbf{3}} - \frac{\boldsymbol{x}}{\boldsymbol{L_i}}\right) \tag{13}$$

where $L_i$ is the length of strip *i*.

The third architecture is based on the concept that the load is resisted by a combination of flexural stiffness effects and prestress. In this case the relationship between the maximum slope, the prestress of strip i and the flexural stiffness is:

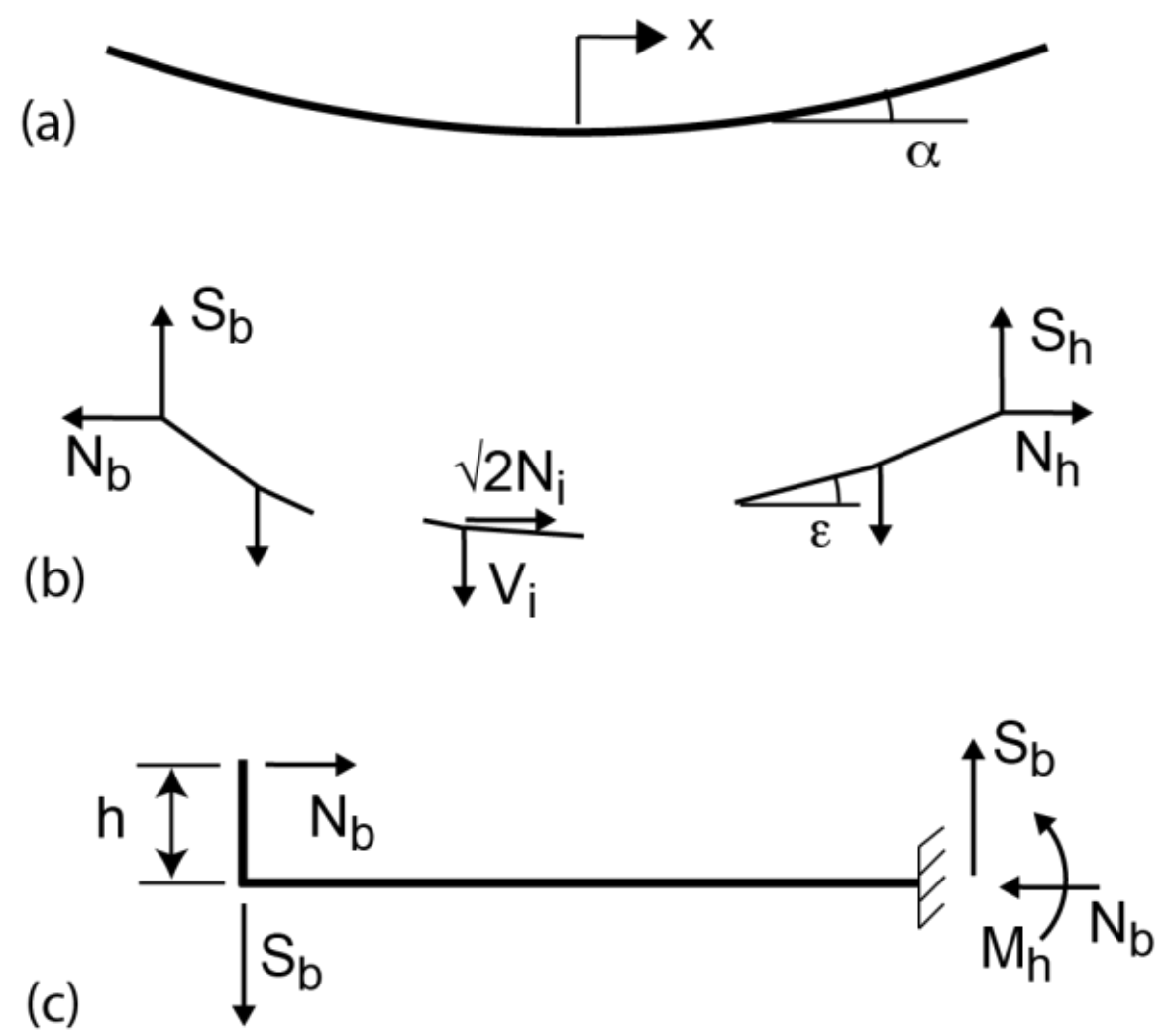

Figure 21: Loads on (a) strip, (b) diagonal cord and (c) boom.

$$\boldsymbol{\alpha} = \frac{\boldsymbol{P\,w}}{\boldsymbol{N_i^{\frac{2}{3}}\,EI^{\frac{1}{3}}}}\,\mathbf{tanh}\left(\sqrt{\frac{\boldsymbol{N_i}}{\boldsymbol{EI}}}\frac{\boldsymbol{L_i}}{\mathbf{2}}\right) \quad - \frac{\boldsymbol{L_i}}{\mathbf{2}\,\boldsymbol{N_i}} \tag{14}$$

For all three architectures, the force at the boom tip is given by:

$$\boldsymbol{N_b} = \sqrt{\mathbf{2}}\,\sum_{\boldsymbol{i}=\mathbf{1}}^{\boldsymbol{n}} \boldsymbol{N_i} + \boldsymbol{N_h} \tag{15}$$

where $N_i = 0$ for Architecture 2.

Figures 22 and 23 show plots of the minimum tip cord tension required to achieve different values of the maximum slopes $\alpha$ and $\varepsilon$, for a spacecraft module with side length $L$ = 60 m and with strips of width $w$ = 1.5 m. In Fig. 22 note that a specific flexural architecture (Architecture 2) with $EI = 10$ Nm$^2$ has been considered, for which the specific value $\alpha = 0.7°$ has been obtained. Also note that the plot for the prestress + flexural architecture (Architecture 3), which has been obtained for the same value of $EI$, achieves even smaller values of $\alpha$ by applying pretension forces on the order of 0.2 N.

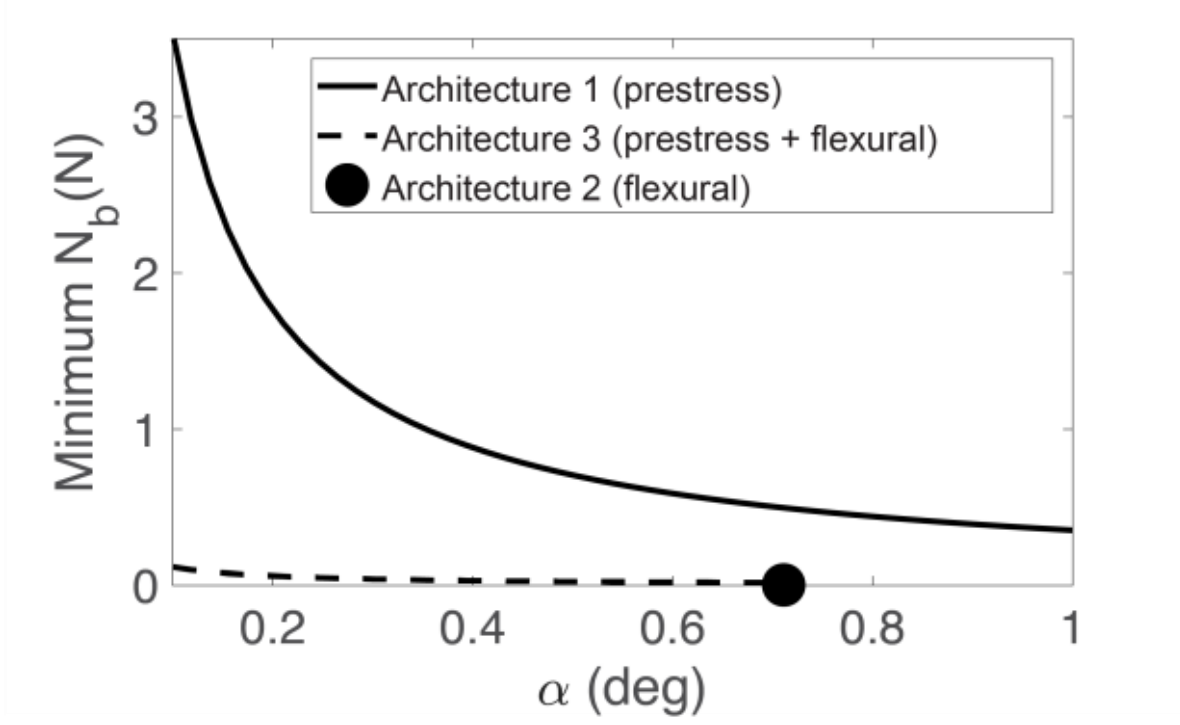

Figure 22: Minimum tip cord tension required for different values of $\boldsymbol{\alpha}$ for three structural architectures, for 60 m x 60 m spacecraft.

In Figure 23 two sizes of diagonal boom have been considered, assuming thin-walled circular tubes with cross-sectional radii of 5 and 10 cm and thickness of 0.5 mm. A tip eccentricity of the cord attachment at the tip of the boom equal to the boom radius has been assumed. Also note that, whereas the trend for $\varepsilon$ is to decrease monotonically for $r$ = 10 cm, for the smaller boom cross-section $\varepsilon$ reverses this trend for larger tension forces. This behavior is due to the cord tension becoming the dominant effect.

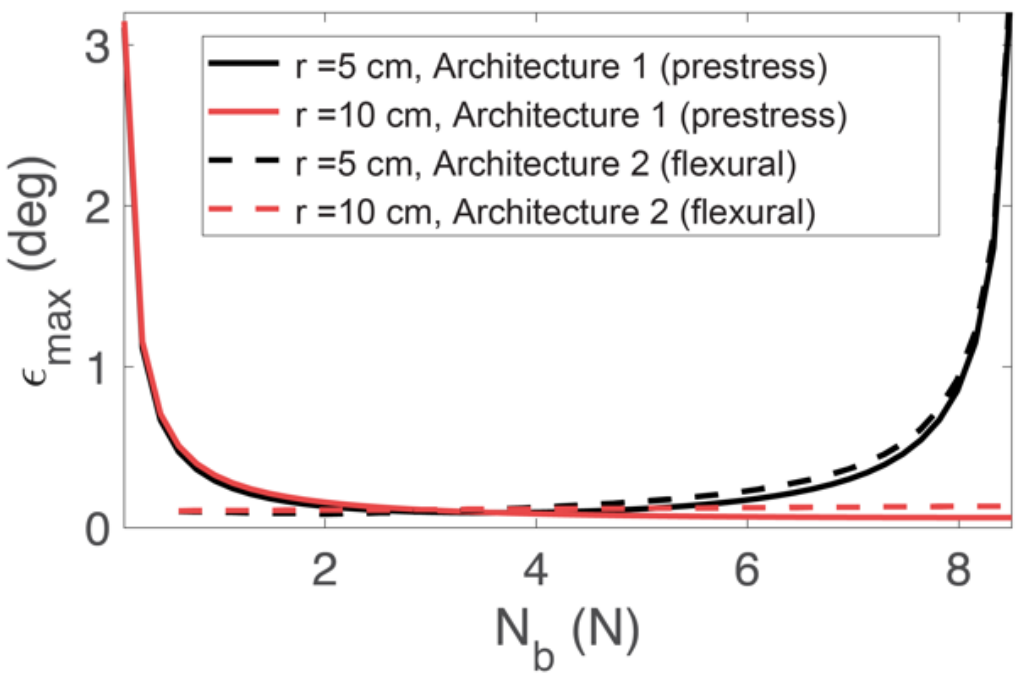

Figure 23: Minimum tip cord tension required for different values of $\boldsymbol{\varepsilon}$, for structural architectures 1 and 2.

This analysis of three structural architectures has shown that the requirement of 1° slope in all directions can be met by all three architectures. Of these, the flexural architecture is the easiest to implement robustly, as it requires the simplest boundary conditions for the strips, and also requires the lowest compressive axial force in the booms. It should be noted that the prestressed architectures are statically indeterminate within the plane of the structure, and it would be very challenging to achieve the desired prestress distributions.

Architecture 2 has been implemented in the strip concept in Figure 24 which shows multiple *battens* connected to two edge *longerons* and supporting the tiles. The longerons run the entire length of the strip and provide the required out-of-plane bending stiffness. To enable the packaging scheme in Section V-A, the longerons must be elastically flattenable and rollable [53]. A cross-section based on the Triangular Rollable and Collapsible (TRAC) boom [54] has been selected.

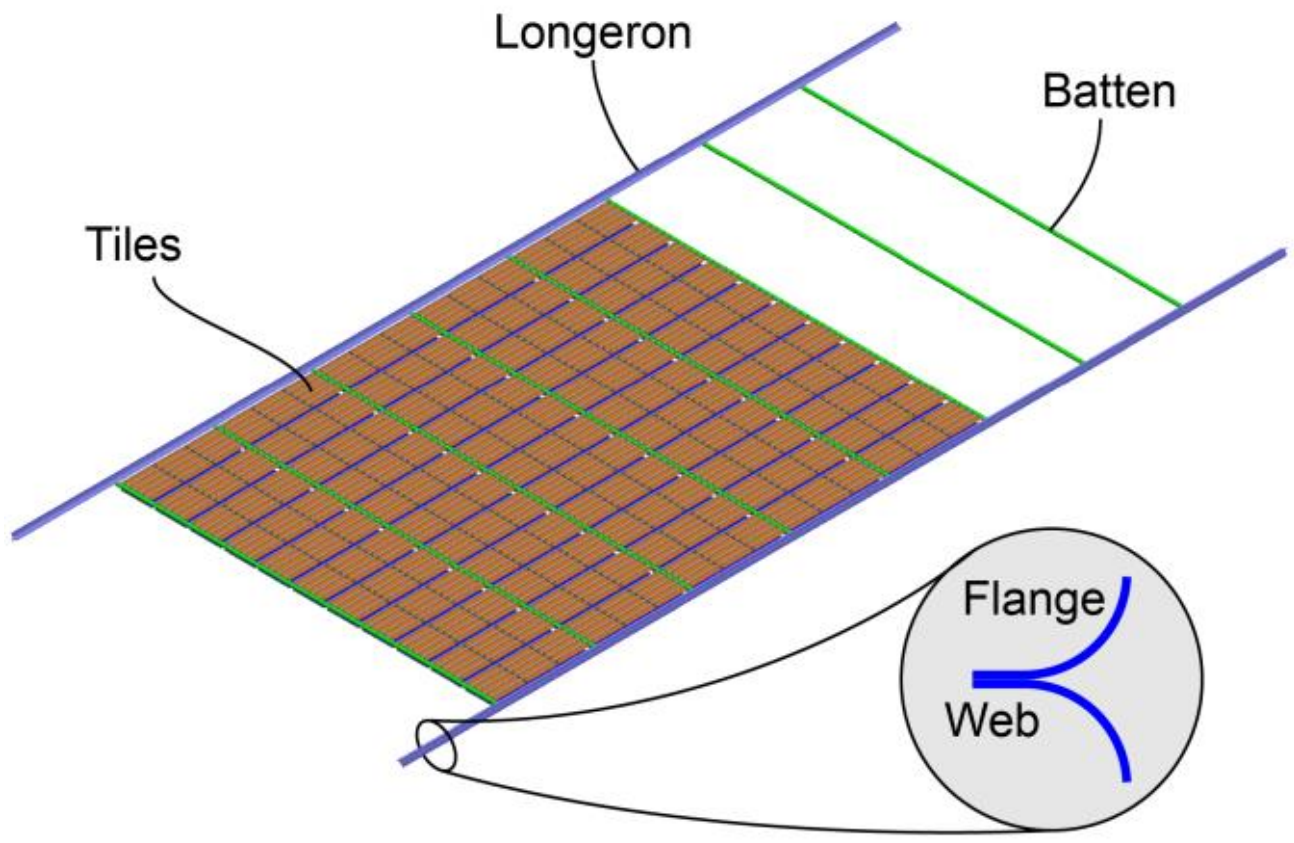

Figure 24: Short segment of a single strip. For clarity, some of the tiles have been omitted.

## *D. Tile Structure*

The tile structure is required to be ultralight, to maintain positional accuracy of the various components, and to flatten for packaging and elastically deploy into its operational configuration. Here we describe the structural design of four tile components: the concentrators, the ground plane layer, the patch antennas layer, and the antenna standoff springs, as shown in Figure 1.

The concentrators consist of a thin, metalized polymer film supported by a deployable, elastic frame structure, with curved edge springs that have the appropriate parabolic profile. Additional springs with identical shape may be included along the length of the concentrator, to achieve a more accurate shape. Each concentrator can be flattened by elastically deforming the parabolic springs.

The ground plane layer consists of a thin metalized polymer film supported along the edges by a thin, square frame. The patch antenna layer is similar in design and is held below the ground plane layer by four springs that have an "S" profile. These springs can be flattened such that the antenna layer rests directly below the ground layer. In the deployed configuration these springs provide the necessary separation between the patch antennas and the ground plane.

Proof-of-concept demonstrators were built at a scale of 10 cm x 10 cm, using 7.5 μm thick polyimide film (Dupont Kapton HN) coated with aluminum for the ground layer and the patch antenna layer, and supported along the edges by frames of 120 μm-thick pultruded carbon fiber rods. The "S" springs were constructed using carbon fiber composite material. The concentrators were made using aluminized 25 μm-thick polyethylene film (Mylar), supported along the edges by carbon fiber composite springs. A pultruded carbon fiber rod was attached to the front surface, along the top edge of the concentrator, and a strip of photovoltaic material was attached to the back surface, along the top edge.

### E. *Module Structural Design*

An analysis of the flexural spacecraft architecture selected in Section V-C, and shown in Figs 20 and 24, was carried out. Following Section V-B it was assumed that the spacecraft is pointed directly at the sun, and the dominant loading case is the solar radiation pressure.

The strips were modeled as beams, the diagonal cords as cables, and the booms as beam-columns. For a spacecraft module with side length $L$ = 60 m, this simplified model has only four structural parameters that control the out-of-plane deflection: the bending stiffness of the booms, the bending stiffness of the strips, the number of strips in each quadrant (which controls the width of each strip), and the diagonal cord tension.

The ATK coilable boom for the ST8 Sailmast was selected. It has a bending stiffness of 15 $kNm^2$ , corresponding to a tip deflection of 55.4 mm and a linear density of 34 g/m [55]. For this choice of boom, we choose the diagonal cord tension $N_b = 3.84$ N corresponding to a maximum slope $\epsilon = 0.05°$ for the diagonal cords, 20 strips per quadrant, and a strip bending stiffness of $EI$ = 10.78 $Nm^2$ corresponding to a maximum strip slope $\alpha = 0.67°$. The Euler buckling load of this boom is ~80 N.

To achieve the desired strip bending stiffness, the two longerons along the strip edges must each have a bending stiffness > 5.4 $Nm^2$. Assuming a longitudinal modulus of 140 GPa for the longerons (a conservative estimate for high-strain carbon fiber composites) and a cross-section with 10mm flange radius and $105°$ subtended angle, a flange thickness of 68.5 μm corresponds to a bending stiffness of 8 $Nm^2$.

The corresponding flattened longeron thickness is 137 μm, and we have assumed that the tiles and battens also have this thickness when they are flattened, in order to estimate the packaged envelope volume. The 60 m x 60 m structure, with 20 strips per quadrant, a flattened strip thickness of 137 μm, and a minimum bend radius of 13.7 mm (corresponding to a maximum longitudinal strain of 0.5% in the longerons), fills 95.6% of a cylinder with a diameter of 0.92 m and height of 1.50 m.

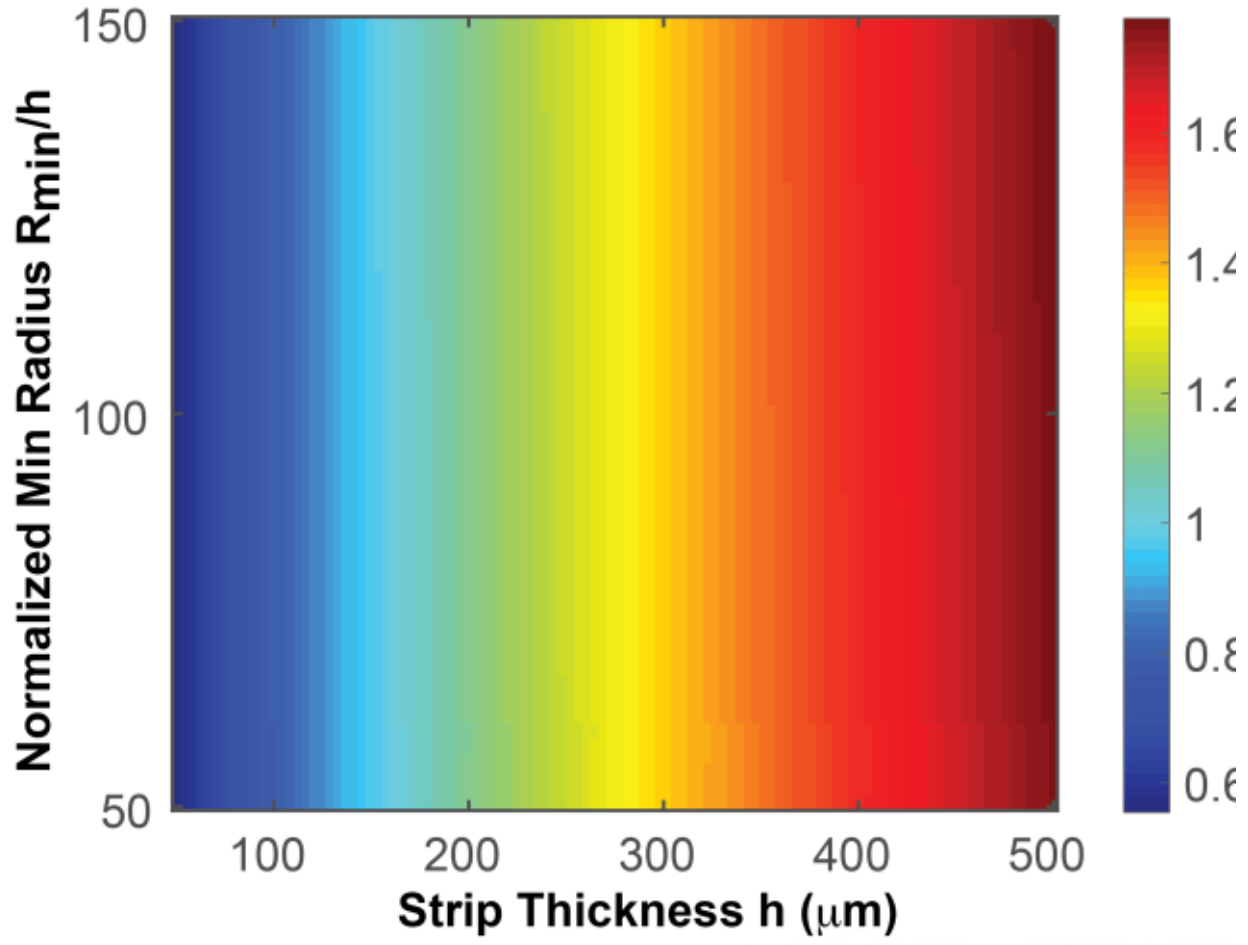

Figure 25. Variation in packaged diameter of 60 m × 60 m module for a range of flattened strip thicknesses (h) and minimum bend radii ($R_{min}$).

Figure 25 shows the effect of varying the flattened strip thickness and minimum bend radius of the strips on the packaged diameter of the spacecraft. For the range 50 - 500 μm of flattened strip thicknesses and the range $R_{min}/h = 50 – 150$ for the non-dimensional minimum bend radius, the packaged module diameter varies between 0.56 m and 1.81 m.

### *F. Total System Mass and Launch Accommodation*

The mass of a 60 m × 60 m module was estimated by accounting for the mass of the tiles, the hub, the strip structure, the booms, and the diagonal cords. The tile mass was calculated by multiplying the expected tile areal density of 160 g/m$^2$, see

| Layer | Material | Thickness (μm) | Mass (g) |
|---|---|---|---|
| Concentrator Reflective | Al | 10 | 0.324 |
| Concentrator Backing | Kapton | 10 | 0.170 |
| Front Emissive Layer | SiO2 | 4 | 0.122 |
| PV Cell | III-V and Cu | 40 | 0.257 |
| Tile Support | Kapton | 10 | 0.142 |
| Routing Layer | Al | 5 | 0.135 |
| Antenna Backing | Kapton | 10 | 0.142 |
| Antenna Conductive | Al | 2 | 0.054 |
| Si IC and Shield | Si/Al2O3 | 300/1000 | 0.116 |
| Carbon Fiber Frame | Carbon | Various | 0.138 |
| **Total** | | | **1.6** |

Table **1** for a detailed breakdown, by the total spacecraft area. The tile mass does not change with changes in the structural design of the spacecraft. The hub mass was assumed to be 50 kg, based on the use of nanosatellite components and including the propulsion system. The mass of the strip structure was calculated by multiplying the cross-sectional area of the longerons by their total length, times the density of CFRP (1600 kg/m$^3$), and the mass of the battens was calculated in a similar way. The diagonal cord mass was calculated by estimating the cord cross-sectional area such that given the desired diagonal cord pre-tension induces a strain of 0.1% and using this area to calculate the cord linear density (using a volumetric density of 1600 kg/m$^3$).

The mass breakdown of this 60 m × 60 m spacecraft, shown in

| Component | Mass (kg) |
|---|---|
| Tiles | 576 |
| Hub | 50 |
| Strip structure (longerons and battens) | 19 |
| Booms | 6 |
| Diagonal cords | 0.01 |
| **Total** | **651** |

Table 2, results in a total mass of 650 kg, corresponding to an overall areal density of 181 g/m$^2$.

Table 1. Breakdown of mass contributions and total mass of a 10 cm x 10 cm tile.

Table 2. Breakdown of mass contributions and total mass of a 60 m × 60 m module.

Our compact packaging technique enables the launch of many modules in a single launch vehicle. Figure 26 shows the number of modules that fit inside the payload fairing of a single launch vehicle for different types of launch vehicles. Except for the NASA Space Launch System (SLS), the number of 0.92 m wide modules that can be accommodated is limited by the total mass limit of the launch vehicle. However, if the packaging efficiency is reduced and the module packaged diameter is increased, we reach a regime where the total number of modules is limited by the available volume.

| Layer | Material | Thickness (µm) | Mass (g) |
|---|---|---|---|
| Concentrator Reflective | Al | 10 | 0.324 |
| Concentrator Backing | Kapton | 10 | 0.170 |
| Front Emissive Layer | $SiO_2$ | 4 | 0.122 |
| PV Cell | III-V and Cu | 40 | 0.257 |
| Tile Support | Kapton | 10 | 0.142 |
| Routing Layer | Al | 5 | 0.135 |
| Antenna Backing | Kapton | 10 | 0.142 |
| Antenna Conductive | Al | 2 | 0.054 |
| Si IC and Shield | $Si/Al_2O_3$ | 300/1000 | 0.116 |
| Carbon Fiber Frame | Carbon | Various | 0.138 |
| **Total** | | | **1.6** |

## G. *Summary*

We have presented a structural design concept for the tile and a preliminary structural design of the 60 m x 60 m module.

The tile structure is designed to be lightweight (with a total density expected to be around 160 g/m$^2$), maintain positional accuracy of the tile components, and elastically flatten for packaging. Initial mockups have demonstrated the ability of the tile to be flattened and return to its original shape.

The module structure consists of lightweight, stiff strips that support the tiles. The whole structure is designed to maintain tile tilts under normal solar pressure loading below 1° while keeping the module mass low. The overall module mass is estimated at 651 kg. It can be elastically packaged into a compact cylindrical form with a diameter of 0.92 m and a height of 1.50 m.

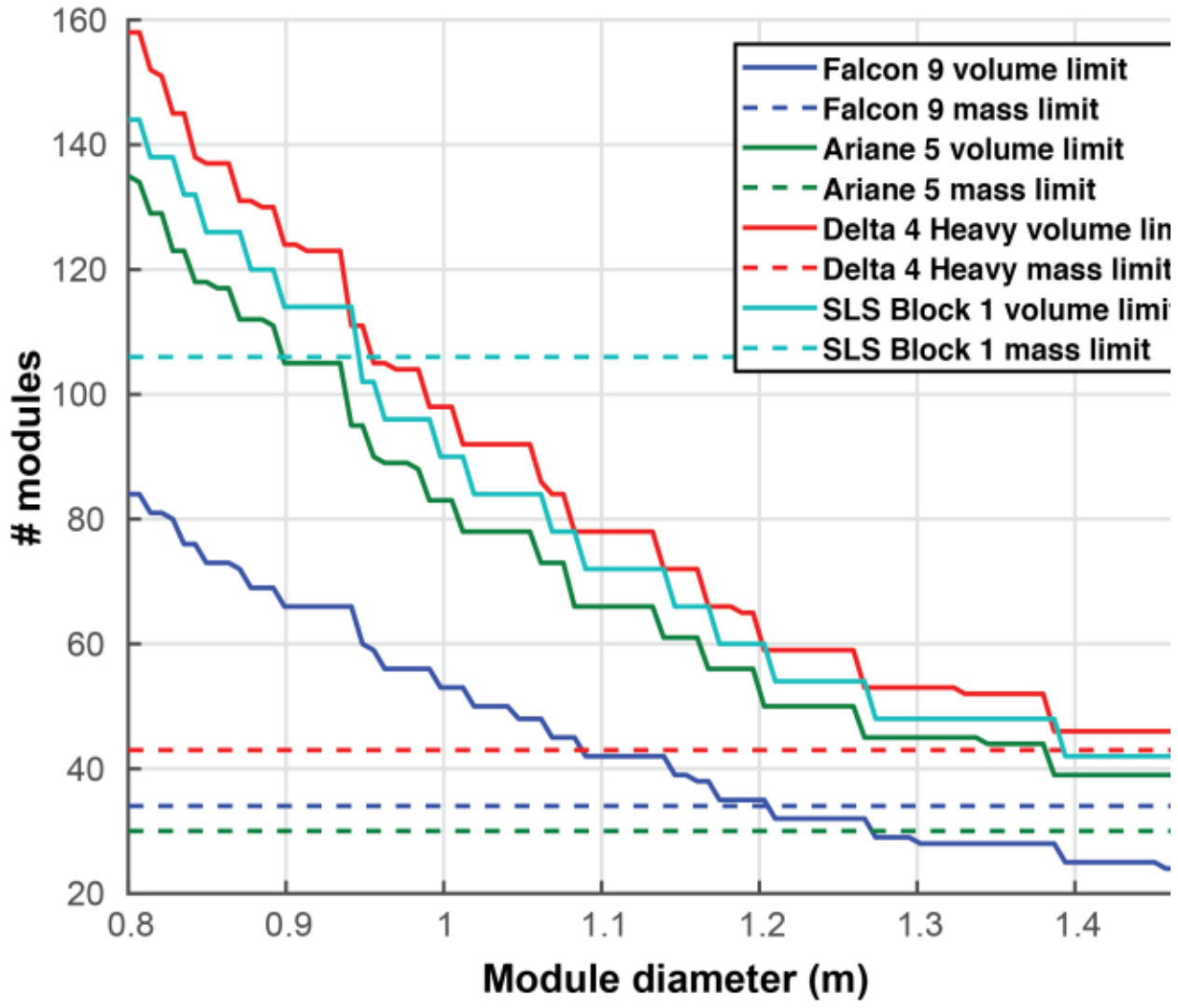

Figure 26. Number of modules that fit within the payload fairing of typical launch vehicles for different diameters of the packaged modules.

# VI. INTEGRATED SYSTEM DISCUSSION

## A. *Introduction*

The size and complexity of a system based on the module approach previously described depends on the amount of power we wish to provide to a terrestrial grid. The power provided to the gird drives the overall size of both the space-based part of the system, the Space Segment (includes launch), and the receiving side of the system, the Ground Segment. Design concepts for the Ground Segment are beyond the scope of this paper but we will account for the various efficiencies encounter when converting the beamed RF power to AC power to the grid. We look at the system parametrically where the power to the grid, $P_g$, is used to develop the system size. We discuss the performance metrics of the module resulting from the subsystems described above. From that, we derive expressions for the mass required on orbit, the number of launches required to get that mass to orbit, the size of the rectenna on the ground, and a very rough estimate of costs and levelized cost of electricity (LCOE) associated with this instantiation of the Space Segment of a space solar power (SSP) system.

## B. *Scaling to The System Level*

As stated previously, the space vehicle – solar power payload and the spacecraft that supports the payload – has an estimated mass of 657 kg (Table V-2). We estimate the RF specific power, S, by:

$$\mathbf{S}(\mathbf{W/kg}) = \frac{\eta_{PV}\eta_{DCRF}\eta_{Tx}}{m_{SV}} A_{pV}(AM0) \qquad (16)$$

where $\eta_{PV}$ = efficiency of the photovoltaics

$\eta_{DCRF}$ = efficiency of converting the DC power into RF power

$\eta_{Tx}$ = transmitting antenna efficiency

$m_{SV}$ = mass of the space vehicle

$A_{PV}$ = effective area of photovoltaic material

AM0 = 1366 W/m$^2$ insolation

The tile described above has goal values of $\eta_{PV}$ = 25%, $\eta_{DCRF}$ = 50%, $\eta_{Tx}$ = 96%, and is incorporated into a module with $A_{PV}$ = 3600 m$^2$ (60m x 60m) yielding a specific power (radiated RF power) of 890 W/kg. The total RF power generated by one space vehicle is $P_{rf}$ = 584 kW.

The SSP system includes a Ground Segment to receive the RF energy, convert it to AC electrical energy and supply that energy to a power grid. This process introduces additional efficiencies:

1. $\eta_{diff}$ which accounts for the energy only in the main lobe of the antenna beam from the SSP satellite(s),
2. $\eta_{RFDC}$ which accounts for the efficiency of the rectifying antennas, or rectennas [REF] in receiving and converting the beamed RF energy to DC electrical energy, and
3. $\eta_{DCAC}$ which is the efficiency in converting the DC energy to AC electrical energy supplied to the grid. We account for internal ohmic and reaction losses in $\eta_{DCAC}$.

In what follows, we take $\eta_{diff}$ = 84%, $\eta_{RFDC}$ = 82% and $\eta_{DCAC}$ = 90% based on current technologies without going into a top-level design for the ground. We assume the area or the rectenna array on the ground is equal to the projected area of the main lobe of the transmitted RF power.

The link budget for the SSP system is modeled using a modification of Friis formula [27],

$$\boldsymbol{P_g = [\eta_{PV}\eta_{DCRF}\eta_{Tx}A_{PV}(AM0)]\left(\frac{f}{cr}\right)^2 A_{PV}A_r\eta_{dif}\eta_{RFDC}\eta_{DCAC}} \quad (17)$$

where f = frequency of operation, $A_{pv}$ = effective area of the photovoltaics = area of the transmitting aperture, r = range from the satellite(s) to the ground reception point, and c= speed of light in vacuum. The quantity in the square brackets is the power transmitted, $P_T$, in the Friis formula. Cleaning up the above equation by identifying $H_s = \eta_{PV}\,\eta_{DCRF}\,\eta_{Tx}$ and $H_g = \eta_{diff}\,\eta_{RFDC}\,\eta_{DCAC}$ we get the interesting result that the received power is proportional to the square of the transmitting antenna area:

$$\boldsymbol{P_g = H_s H_g \left(\frac{f}{cr}\right)^2 A_{PV}^2 A_r (AM0)} \quad (18)$$

The most efficient use of the SSP beamed power comes from sizing the rectenna array to fill the projected area of the main lobe from the SSP satellite(s). Depending on the orbit and the size of the transmitting aperture, the receiving rectennas can be either in the near or far field. We model $A_r$ for both conditions as follows [56]:

$$\boldsymbol{A_r = (r\theta)^2 = r^2\begin{cases}\frac{A_{PV}}{r^2}+\frac{c^2}{f^2A_{PV}} & r < 2A_{PV}f/c \\ \frac{c^2}{f^2A_{PV}} & r > 2A_{PV}f/c\end{cases}} \quad (19)$$

Combining equations 17 and 19 results in a link budget in which only the transmitting antenna area appears:

$$P_r = \begin{cases} \left[\left(\frac{fA_{PV}}{cr}\right)^2 + 1\right] H_s H_g (AM0) A_{PV} & r < 2A_{PV} f/c \\ H_s H_g (AM0) A_{PV} & r > 2A_{PV} f/c \end{cases} \qquad (20)$$

With the help of equation 20, the previously described the mass and specific power estimates for the space vehicle, and the overall efficiencies of the Ground Segment, we can estimate the number of space vehicles needed to provide $P_g$ to a power grid, and the number of launches required to achieve the number of vehicles on orbit.

Our initial concept is for a formation flying group of space vehicles in GEO capable of beaming power to the Earth from ~ 60° S latitude to 60° N latitude for the hemisphere under the SSP space system. We take the range to the ground station as 40,000 km.

For a given $P_g$, the number of vehicles on orbit $N_{SV}$ is:

$$N_{SV} = \left\lceil \frac{P_g}{H_s H_g A_{PV} (AM0)} \right\rceil \qquad (21)$$

Note the ceiling operation indicated by the brackets.

There are a limited number of launchers at present, capable of putting large payloads into GEO [57]. We characterize a given launch system's capability by $M_{GEO}$, the mass that the launcher can place into GEO, and κ, a de-rating factor for the launcher, acknowledging that some reserve mass is needed to accommodate uncertainty in the mass of the space vehicles and supporting structure for the space vehicles on the launcher [58]. Given κ and $M_{GEO}$, and the mass of the space vehicle, we can compute the number of space vehicles that a given launcher can place into GEO, $n_{SV}$:

$$n_{SV} = \left\lfloor \frac{\kappa M_{GEO}}{m_{SV}} \right\rfloor \quad 0 < \kappa < 1 \qquad (22)$$

The number of launches needed to place all $N_{sv}$ space vehicles into GEO is simply:

$$N_L = \left\lceil \frac{N_{SV}}{n_{SV}} \right\rceil \qquad (23)$$

Note the ceiling brackets – this acknowledges that you cannot launch a fraction of space vehicle and must use another launcher.

With knowledge of Nsv and $N_L$, we begin to see how the system cost, and ultimately the levelized cost of electricity scale with $P_g$, S, and $m_{SV}$. Figure 27 shows the trends for the number of space vehicles and launches as a function of power to the grid based on a Falcon 9 Heavy launcher capable of placing 3,000 kg (3 MT) to GEO. Since both $N_{SV}$ and $N_L$ are linearly dependent on $P_g$, the details of the space vehicle (mass, area, specific power) determine the slopes of the graphs of both quantities as function of $P_g$.

of the four classes. We assume that the SSP space vehicles are similar to Class A missions and use $200,000/kg as a cost scaling figure. Based on the estimated 369 kg mass for the space vehicle, the estimated cost is $96M per vehicle.

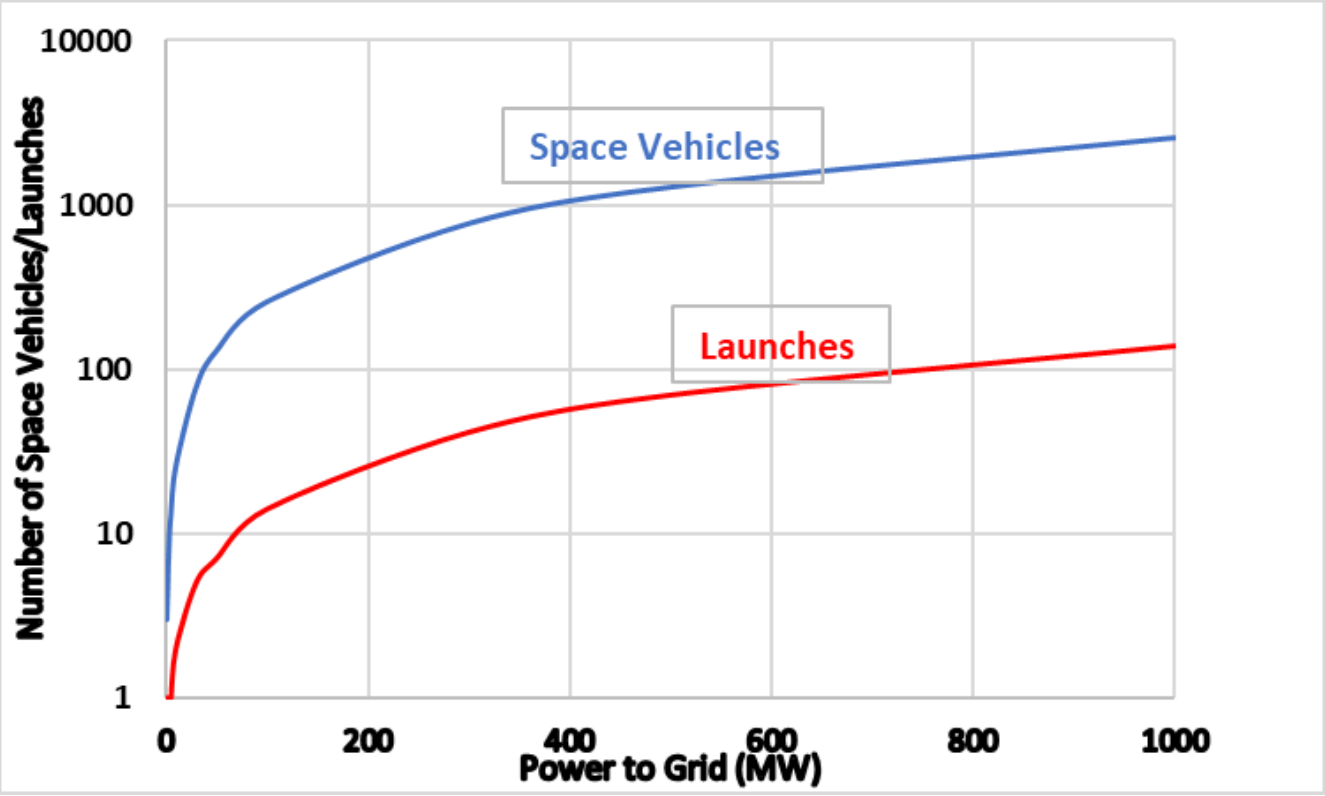

Figure 27. $N_{SV}$ and $N_L$ are linearly dependent on $P_g$ – the slopes of the curves depend on the specifics of the space vehicle.

Based on these relationships, we can estimate the cost of the space portion of an SSP system.

To get a reasonable estimate for the space vehicle cost (payload, spacecraft, integration, testing, etc.), $C_{SV}$, we employ a rule of thumb that space vehicles cost between $90,000 - $250,000/kg on -orbit. The variation has to do with the space vehicle complexity, mission requirements, operational lifetime, etc. Another way of looking at the cost spread is the difference between building satellites for Class A, B, C, or D missions [59]. Class A is an operational mission with lowest possible risk, and is usually the most costly of the four classes. We assume that the SSP space vehicles are similar to Class A missions and use $200,000/kg as a cost scaling figure. Based on the estimated 369 kg mass for the space vehicle, the estimated cost is $96M per vehicle.

Launchers cost ($C_L$) between $90M for a Falcon 9 Heavy to $200M for an Ariane 6. A reasonable upper limit for the space portion cost is:

$$\boldsymbol{C_{SS} = C_{SV} \times N_{SV} + C_L(\$M)N_L} \quad (24)$$

Equation 24 ignores any cost decrease due to learning curves to produce the satellites and the launchers. Learning curves acknowledge the increase in efficiency of production of an item as the number of items produced is increased [60]. The equation also assumes every launch is successful and that every satellite works to specification once on orbit. We acknowledge that these are optimistic assumptions and, in that light, equation 24 provides a lower upper bound on cost for the space segment given the assumptions above.

The estimate of LCOE for the space segments requires an assumption about the space segment lifetime, T. Assuming the satellites operate at GEO, then they can produce energy 365 days a year, 24 hours a day except for twice a year during equinoxes. The total yearly eclipse time is approximately 80 hrs (The maximum daily eclipse time is less than 1 hour 10 minutes) so the total solar illumination time is about 8,684 hrs. per year. The total energy delivered to the ground, $E_g$, is:

$$\boldsymbol{E_g = 8{,}684 P_g(kW) \times T(yrs) \quad kW - h} \quad (25)$$

where $E_g$ is in kW-hours. The LCOE contribution for the space segment ($LCOE_{ss}$) is computed by:

$$\boldsymbol{LCOE_{SS} = \frac{C_{SS}}{E_g} \quad \$/kW - h} \quad 26$$

shows what the space segment requires and the LCOEss for satellites providing 50 MW of power to the grid from GEO.

While the predicted LCOE contribution from the space segment is high compared to terrestrial electricity prices, the exercise in getting to this number provides valuable insight as to what directions research should take. First, mass is everything. The cost model assumes that cost scales with mass and this has generally held true for space systems. Thus, further efforts in reducing the overall mass can have a large impact on the overall economic feasibility of the system.

Second, efficiencies are important. There are a few areas where efficiencies may be able to be increased. The photovoltaic efficiency could be driven higher by pursuing research and development of new PV materials. DC to RF
conversion efficiencies could be improved with an updated integrate d circuit design, or by looking at Si-Ge fabrication processes for better X band efficiencies. There are several places in the ground segment where efficiencies may be increased as well, notably in the rectenna.

| | | |
|---|---|---|
| | | |
| | | |
| | | |
| | | |
| | | |
| | | |
| | | |

The above analysis also assumed the space vehicles can convert solar energy to RF energy over the entire day with exception of eclipse season. However, the planar geometry of our system restricts the time we can provide energy to the ground as the RF face of the system points out into deep space as the system goes into local midnight ( Figure 28). This cuts the energy production effectively in half, thereby doubling the LCOE. The next section discusses this further.

Table 3. The space segment contribution to LCOE is very high relative to current terrestrial prices. This is driven by the number of space vehicles required to meet the 50MW requirement.

### C. *Orbital Considerations*

One of the primary challenges with space-based solar power is the design and maintenance of a satellite constellation in formation flight. In the past decade, substantial amount of research has been done on the guidance, navigation and control of formation flying satellites [61], [62], [63], [64]. Formation flight has also been demonstrated in space by missions such as GRACE [65], GRAIL [66], TanDEM-X [67] and PRISMA [68]. The detailed design of the formation flying constellation along with the associated sensing and actuation requirements is currently being pursued and will be presented at a later date.

| **Quantity** | **Value** | **Comments** |
|---|---|---|
| $N_{sv}$ | 130 | |
| Total Sat. Cost | $17.39B | |
| $N_L$ | 7 | Falcon 9 Heavy |
| $C_L$ | $0.9 | |
| Space cost | $17.84B | |
| $LCOE_{ss}$ | $2.05/kW-h | 20 years on-orbit life |

A critical decision in the orbital design of the space solar constellation is the choice of the orbit altitude. From a launch cost perspective, low Earth orbits (LEO) are easier to get to and place less stringent requirements on the beamwidth of the antenna array. But a LEO constellation would not be able to generate power ~40% of the time on account of being eclipsed by the Earth. While the constellation could be placed in a terminator orbit (polar sun-synchronous 6am-6pm), it leads to a highly inefficient orientation for RF transmission. The LEO constellation would also require a network of ground-based receivers on Earth to continuously relay power from the space-based array. Moreover, in low Earth orbit, one must deal with orbital perturbations due to atmospheric drag, Earth oblateness (J2) and solar radiation pressure, further complicating the guidance, navigation and control problem.

On the other hand, a constellation in GEO can radiate all its power to a single ground-based receiver. The spacecraft are always in view of the sun, except for a few days in the year close to the equinoxes when the Earth eclipses the sun for up to an hour each day. While the electronics in GEO must survive a harsher radiation environment than LEO, maintaining a constellation in formation flight is relatively easier since we only have orbital perturbations from solar radiation pressure. Keeping these factors in mind, the point design presented in this paper assumes that the constellation is in GEO.

*D. Capacity Factor*

The total power delivered to the receiver array on Earth depends on two major geometrical factors 1) angle made by the photovoltaic concentrators with the incoming solar radiation 2) angle made by the transmitting patch antenna array with the receiving station on Earth. Taking these two factors into account, we can estimate the desired optimum orientation for the modules at each location in the geostationary orbit. This optimum orientation depends on whether the RF energy can be transmitted through the photovoltaic layer and these two scenarios are presented in Figure 28.

As shown in the figure, the modules rotate continuously to achieve the right balance between collecting solar power at a favorable angle and transmitting RF power efficiently to the ground receiver. In the case of dual-sided transmission, power is generated and transmitted throughout the orbit but there is a rather sudden flip in the orientation, when the modules are at locations corresponding to 6 am and 6 pm local time. The single sided transmission case does not require any sharp attitude maneuvers, but the system has to go through a phase of not being able to transmit any power to the ground receiver.

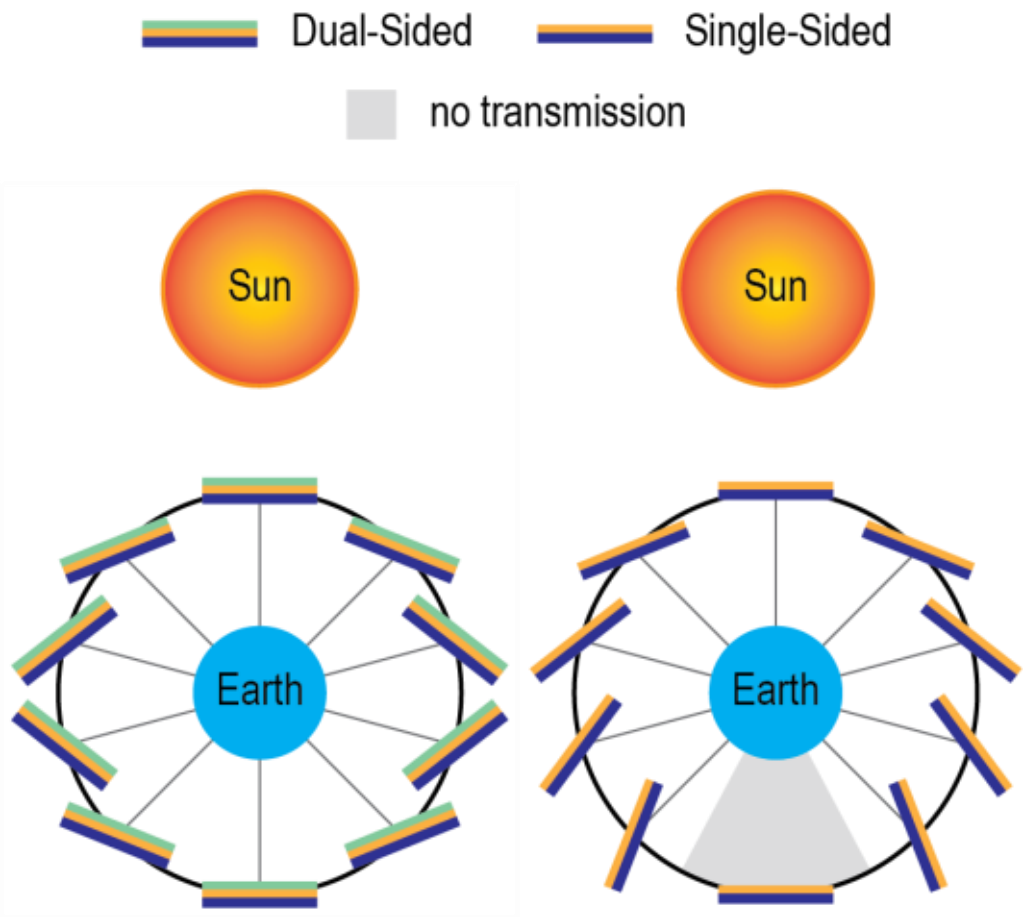

Figure 28. Single and dual-sided capacity factor schematic.

Based on current estimates, the dual-sided system is expected to deliver 1.56 times more power than the single-sided system. However, this would require the design of RF-transparent photovoltaic concentrators with two separate planes for RF transmission or equivalently, PV-transparent RF antennas. Engineering solutions for dual-sided systems and their corresponding mass penalties are currently being evaluated. In this paper, the single-sided module is chosen as the baseline for the design of the structural, photovoltaic and RF sub-systems.

*E. System Considerations for Environmental Radiation*

We evaluated the radiation environment for our space-based solar power system operating in GEO and LEO orbits including the effect of trapped protons and electrons, solar protons and galactic cosmic rays. Total Ionization Doses (TID) and Displacements Damage Dose (DDD) are presented for different space-based solar power components that we have described in the current paper. Results are summarized to evaluate potential system degradation and to perform system trade studies, optimizing main component design for radiation hardness, radiation shielding, and expected power profile over the system operating lifetime. The initial focus is on ionizing effects on micro-electronics and concentrating mirrors degradation (TID analysis), and non-ionizing radiation effects on solar cells degradation (DDD analysis); evaluation of surface charging and internal charging analysis will be performed in the future.

For initial studies with preliminary shielding geometries, propagation of the radiation environment through the tile (depicted in Figure 1) instrument was modeled using NOVICE code, which is a three-dimensional adjoint (reverse) Monte Carlo (MC) transport simulation [69]. The incident electron and proton spectrums are from the AE9 and AP9 models [70] applied to 15 years operational time.

The geostationary orbit is in the outer radiation belt and is dominated by high energy trapped electrons and solar protons from the solar events, and these particles are the primary sources of solar cell degradation. For typical spacecraft solar panels in GEO orbit with ~75-100 μm of the front coverglass shielding and semi-infinite back shielding as a solid back panel, solar cells are exposed to $\sim10^{14}$ $e^-/cm^2$ equivalent 1 MeV electron fluence after 15 years of operation [71] [72]. Based on simulations and several on-orbit solar arrays satellite telemetry measurements over extended period of time, this exposure causes maximum degradation of ~87-90% BOL performance for multi-junction (MJ) GaAs solar cells depending on details of the cell technology and operational period with respect to the solar cycle [71] [73]. Figure 29 shows the simulated results for 1 MeV equivalent electron fluence experienced by SSPI solar cells in the proposed ultra-light CPV design as a function of coverglass thicknesses for 15 years operation at GEO orbit.

For MJ GaAs based solar cell and the SSPI concentrating photovoltaic design (CPV) with 75 μm of front coverglass and mass equivalent shielding on the reverse, 1 MeV equivalent electron fluence at the cell interface are at higher level compared to a typical system due to ultra-light design: $\sim10^{15}$ $e^-/cm^2$ after 15 years of GEO operation. The mass of the coverglass is a major factor in the overall mass of the CPV subsystem. This presents the challenge of achieving SSPI CPV End Of Life (EOL) performance in the GEO radiation environment, while maintaining a physically light and flexible system design. The above parameters are chosen as an initial SSPI system shielding corresponding to ~80% of beginning of life (BOL) power after 15 years of GEO operation [74] [75]. Advanced technologies in two critical areas are under consideration which will be assessed in future trade studies: 1. Coverglass development, and 2. radiation hard ultrathin solar cells. Different thin solar cells technologies are being considered to use in the SSPI system such as MJ inverted metamorphic (IMM) solar cells and MJ latticed matched ELO cells. Recent cell design developments have already demonstrated significant improvement at the EOL performance of thin solar cell technologies with compatible radiation hardness to current MJ lattice matched (GaInP/GaAs/Ge) solar cells. Experimental results for 1 MeV electrons at a fluence of ~$10^{15}$ $e^-/cm^2$ irradiation have been reported for different technologies: SolAero (Emcore) has reported for 4J IMM cell a remaining EOL power of 82% [76]. SHARP and JAXA collaboration [77] show remaining power factor of 84% to 86% for 3J IMM. To optimize SSPI collected power profile over lifetime of operation, solar cell radiation tests of several generations of thin cells will be required to accurately predict degradation, using SSPI specific cells geometry and different types of protective coverglass material. These simulations and tests could potentially reduce the thickness of the coverglass used in the current design.

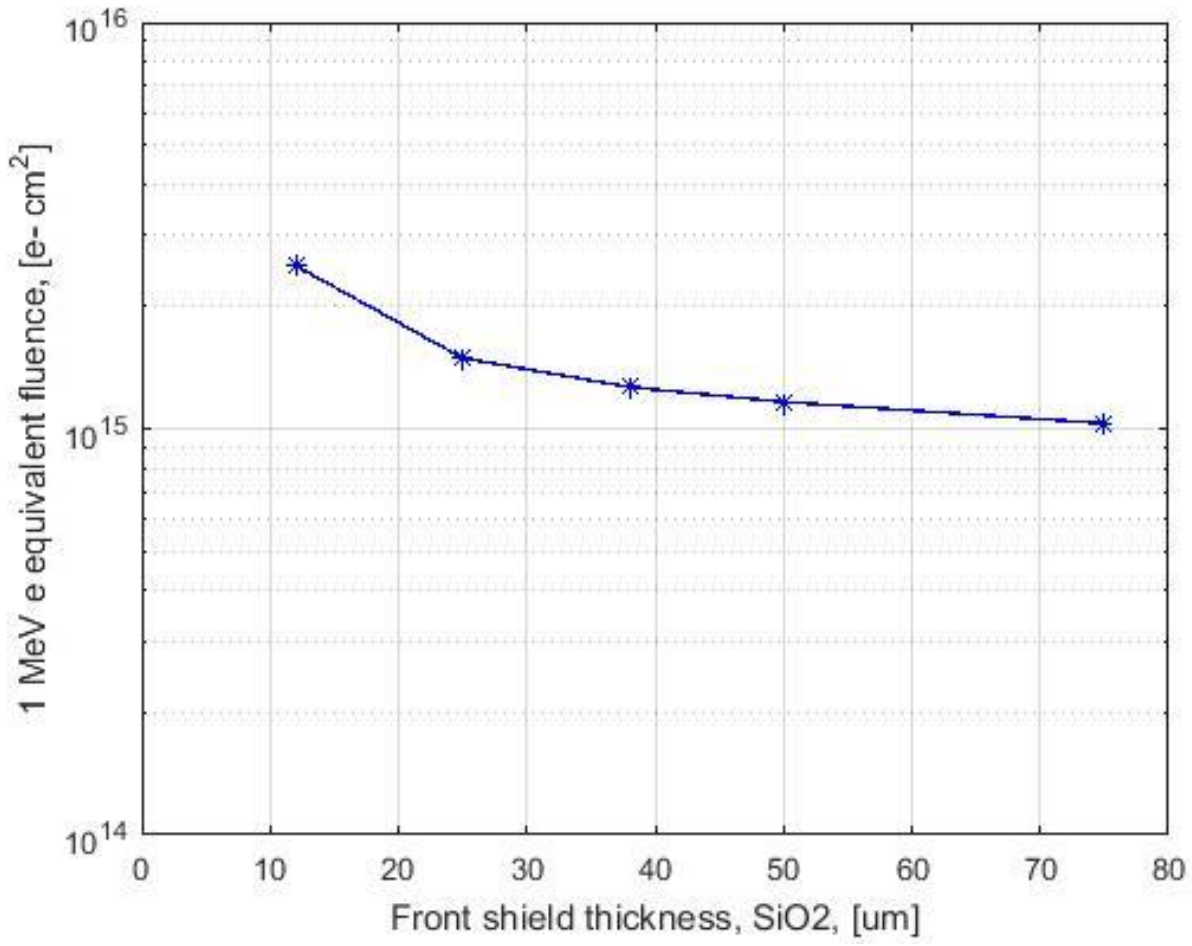

Figure 29 : Simulated 1 MeV e- equivalent fluence experienced by the solar cells in SSPI design as a function of coverglass thicknesses.

For the SSPI integrated circuit (IC), the trapped electrons flux is a dominating factor for TID in the GEO environment. NOVICE simulated IC depth / dose curves show the IC will have an absorbed dose of 1Mrad with ~ 40 mils (1000 microns) of ceramic shielding for 15 years of operation. (Figure 30).

This amount of shielding is not a challenge for system weight considerations due to the small size of the IC. To mitigate radiation effects at these levels, there are several novel approaches to fabricate rad-hard components, even in commercial production, by applying advanced design techniques [78].

At LEO, with opportunities for initial SSPI scaled demonstration, it is becoming common to use commercial-off-the-shelf (COTS) components. Low Earth Orbit spans a range of altitudes which starts below, and reaches into, the inner Van Allen Belt. Below the inner Van Allen radiation belt (approximately 600 km) the environment is relatively benign, and the low energy electron environment is orders of magnitude lower than at GEO orbit The TID environment at LEO increases steeply with altitude as the orbit enters the inner Van Allen Belt which has high concentrations of both trapped electrons and trapped protons. Specific consideration will be required for any particular proposed LEO orbit testing for SSPI small scale operation for concept demonstration.

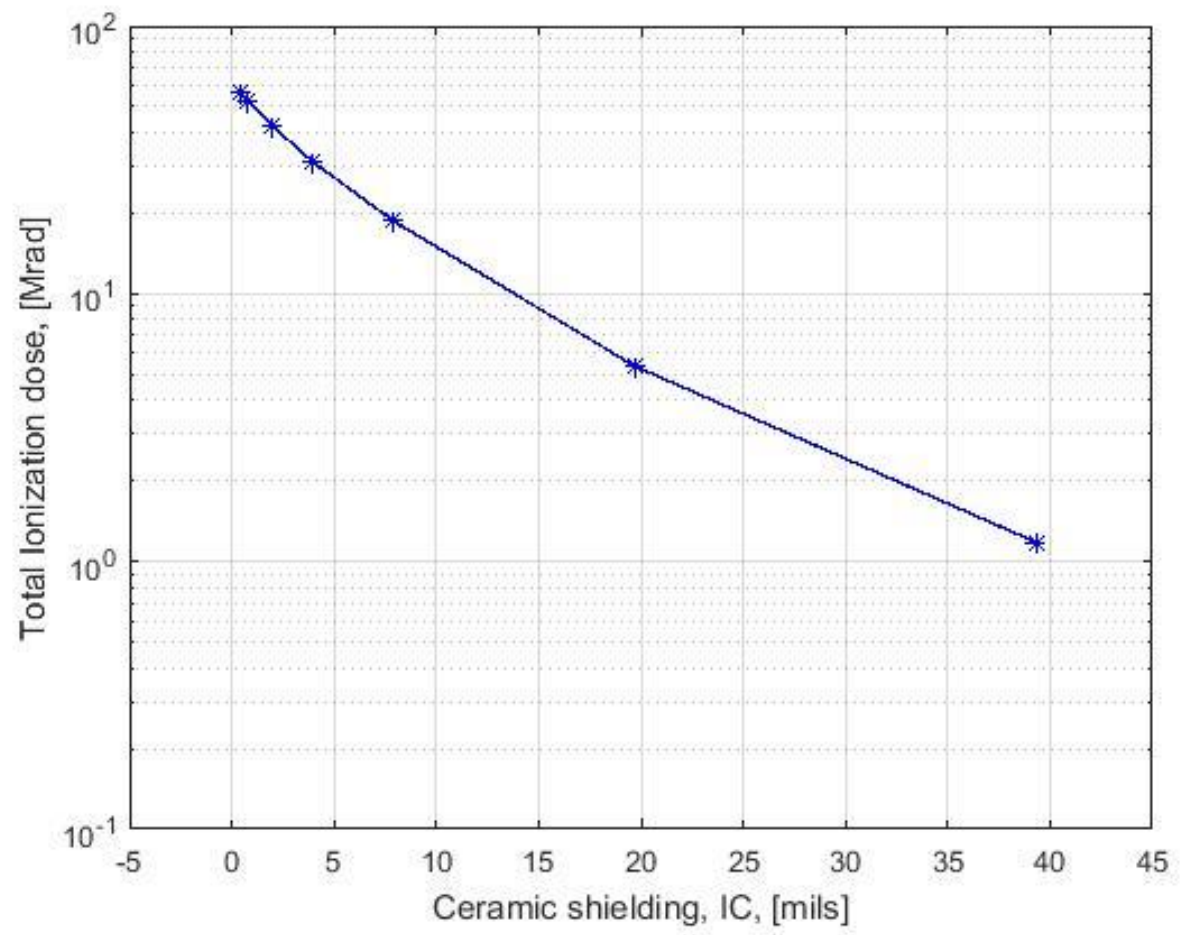

Figure 30: Simulated Total Ionizing Dose Versus Ceramic Shield Thickness for SSPI IC unit, 15 years operation in GEO.

*F. Ground Receiver and Rectenna*

The ground receiver is designed as an array of RF to DC converters (rectennas) that collect and convert the transmitted RF power on Earth. A general structure for a rectenna element consists of the receive antenna, a low pass filter, diodes for RF-DC conversion and a filtering element to suppress AC components in the rectified waveform (Figure 31).

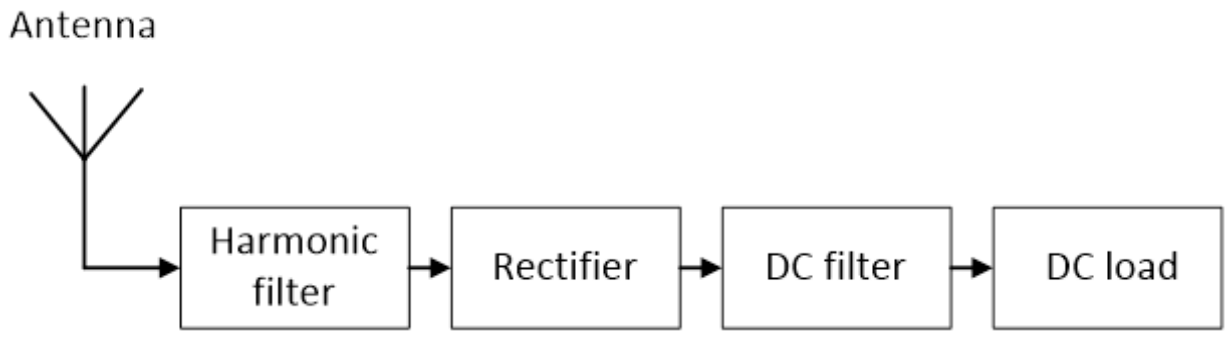

Figure 31 : General rectenna structure.

One of the most important performance metrics of the ground rectennas is the RF-to-DC power conversion efficiency. Besides losses in the passive impedance matching and filtering components, various non-idealities in the diode, such as ohmic resistance, non-zero turn-on voltage and package inductance and junction capacitance, limit the conversion efficiency and maximum frequency of operation. Based on results reported in the literature, we expected conversion efficiencies in the range of 60%-85% at frequencies of interest using current technologies as evidenced by recently reported efficiencies of 73%, 85%, 83%, and 60% at 2.45GHz [79], 2.14GHz [80], 4.5GHz [81] and 10GHz [82], respectively.

*G. RF Safety Discussion*

RF radiation is most often defined as electro-magnetic radiation at frequencies of 3 KHz – 300 GHz. RF radiation is non-ionizing, as the associated photon energy, given by the Planck-Einstein relation $E = hf$ with $h \approx 4.136 \cdot 10^{-15}\ eV \cdot s$, is less than 1.25 meV and thus not high enough to ionize common atoms. Many studies have investigated the effect of RF radiation on living organism or biological tissue. In 1992, a large study by Chou et al. [83] investigated the potential effects of long-Term microwave irradiation on rats, by exposing them to 0.4-W/kg SAR at 2450MHz for 13 months. It followed various parameters, including behavior, blood chemistry/hematology, metabolism, and total body analysis, but found no definitive biological effect in rats chronically exposed to RF radiation at those frequencies. Lai and Singh [84] showed increased amount of single and double stranded DNA breaks in rats exposed to radiation resulting whole body SAR of 1.2W/kg. However, an attempt by Malyapa et al. to measure similar results in alkaline comet assay under similar SAR rates while maintaining constant assay temperature, resulted no significant difference from the control group [85]. In 2010, the INTERPHONE study group found no link between cellular phone usage [86]. The study observed no increased risk of glioma or meningioma to the average user and is referred here since cellular phones operate expose humans to RF frequencies and power levels of interest. As a result of a multitude of these kinds of study, the FCC has concluded (e.g. [87]) that it is currently unknown whether health hazards exist due to non-thermal effects of RF radiation as no such hazards have been conclusively shown.

This has led regulators to adopt guidelines for maximum recommended RF powers density exposure, which are derived from RF thermal effects on human body. For example – the American FCC recommends that professionals will be exposed to RF radiation with density of less than $50 W/m^2$ averaged over 30 minutes, an amount similar to the power levels encountered in the main lobe on the ground. Additional measures such as protective clothing for personnel operating or maintaining the ground station can be adopted, if such personnel is required to be present during live operation.

The end to end efficiency of the whole system is estimated using the individual efficiencies calculated in the previous sections. Taking a photovoltaic efficiency of 25-30%, a DC to RF conversion efficiency of 55-65%, a transmission/collection efficiency of 84%, and a rectenna efficiency of 60-85% we obtain a end to end efficiency of 7-14%.

## VII. Conclusion

We have presented the initial design phase for a lightweight, high-performance space-based solar power array for operation in geosynchronous orbit with an areal mass density of 160 g/m$^2$ and an end-to-end efficiency of 7-14%. A parabolic concentrator made of lightweight polyimide and supported by carbon fiber springs focuses sunlight at a concentration of 20 suns on a multijunction solar cell with an efficiency of ~30%. Excess heat is conducted away from the photovoltaics by a thin film of aluminum and radiated away into space via high emissivity thermal surfaces coating the polyimide support. The converted DC power is routed to a Si integrated circuit where an AC signal is synthesized and amplified at 1-10GHz. This entire process is done and contained entirely in a single, compact, foldable unit cell containing all the necessary elements. Power is then radiated to Earth by microwave antennas across an array of tiles, fully synchronized and phase controlled with Si integrated circuit technology. Arrays of power generating tiles are combined to form free flying modules, which we show are able to be efficiently packaged, stowed and deployed. Modules can be combined to form arbitrarily large space-based power stations. The microwave radiation is collected on Earth with ground-based rectenna stations. Because of the frequency used and the size of our satellite, the incident microwave intensity on Earth never exceeds the intensity of common sunlight and should not be harmful to terrestrial life. The design presented in this paper achieves significant weight reduction and improved efficiency over all previous space-based solar power designs and can be easily tested without full scale deployment. Our proposal is a significant step towards realization of the concept of space-based solar power, first imagined over 70 years ago.

## VIII. Acknowledgements

This works was supported by Caltech Provost Funds, the Caltech/Northrop Grumman Space Solar Power Initiative, and the Caltech Space Solar Power Project.